\documentclass[aps,pra,floatfix,amsmath,ams,superscriptaddress,showpacs,12pt]{revtex4-1}
\usepackage{graphicx}
\usepackage{amsmath,amssymb}
\usepackage{amsfonts}
\usepackage{mathrsfs}
\usepackage{epsfig}
\usepackage{dcolumn}
\usepackage{enumerate}
\usepackage{amsthm}

\begin{document}

\title{ Entanglement between two subsystems, the Wigner semicircle and extreme value statistics}

\author{Udaysinh T. Bhosale \footnote{e-mail: bhosale@physics.iitm.ac.in}}
\affiliation{Department of Physics, Indian Institute of Technology Madras, Chennai, 600036, India}
\author{Steven Tomsovic \footnote{e-mail: tomsovic@wsu.edu}}
\affiliation{Department of Physics and Astronomy, Washington State University, Pullman, Washington, 99164--2814, USA}
\author{Arul Lakshminarayan \footnote{e-mail: arul@physics.iitm.ac.in}}
\affiliation{Department of Physics, Indian Institute of Technology Madras, Chennai, 600036, India}

\preprint{IITM/PH/TH/2010/11}
\begin{abstract}
 The entanglement between two arbitrary subsystems of random pure states is studied via properties of the density matrix's partial transpose, $\rho_{12}^{T_2}$.  The density of states of $\rho_{12}^{T_2}$ is close to the semicircle law when both subsystems have dimensions which are not too small and are of the same order. A simple random matrix model for the partial transpose is found to capture the entanglement properties well, including a transition across a critical dimension.  Log-negativity is used to quantify entanglement between subsystems and analytic formulas for this are derived based on the simple model. The skewness of the eigenvalue density of $\rho_{12}^{T_2}$ is derived analytically, using the average of the third moment over the ensemble of random pure states. The third moment after partial transpose
 is also shown to be related to a generalization of the Kempe invariant.  The smallest eigenvalue after partial transpose is found to follow the extreme value statistics of random matrices, namely the Tracy-Widom distribution. This distribution, with relevant parameters obtained from the model, is found to be useful in calculating the fraction of entangled states at critical dimensions.  These results are tested in a quantum dynamical system of three coupled standard maps, where one finds that if the parameters represent a strongly chaotic system, the results are close to those of random states, although there are some systematic deviations at critical dimensions.

\end{abstract}

\maketitle

\newcommand{\newc}{\newcommand}
\newc{\beq}{\begin{equation}}
\newc{\eeq}{\end{equation}}
\newc{\kt}{\rangle}
\newc{\br}{\langle}
\newc{\beqa}{\begin{eqnarray}}
\newc{\eeqa}{\end{eqnarray}}
\newc{\pr}{\prime}
\newc{\longra}{\longrightarrow}
\newc{\ot}{\otimes}
\newc{\rarrow}{\rightarrow}
\newc{\h}{\hat}
\newc{\bom}{\boldmath}
\newc{\btd}{\bigtriangledown}
\newc{\al}{\alpha}
\newc{\be}{\beta}
\newc{\ld}{\lambda}
\newc{\sg}{\sigma}
\newc{\p}{\psi}
\newc{\eps}{\epsilon}
\newc{\om}{\omega}
\newc{\mb}{\mbox}
\newc{\tm}{\times}
\newc{\hu}{\hat{u}}
\newc{\hv}{\hat{v}}
\newc{\hk}{\hat{K}}
\newc{\ra}{\rightarrow}
\newc{\non}{\nonumber}
\newc{\ul}{\underline}
\newc{\hs}{\hspace}
\newc{\longla}{\longleftarrow}
\newc{\ts}{\textstyle}
\newc{\f}{\frac}
\newc{\df}{\dfrac}
\newc{\ovl}{\overline}
\newc{\bc}{\begin{center}}
\newc{\ec}{\end{center}}
\newc{\dg}{\dagger}
\newc{\prh}{\mbox{PR}_H}
\newc{\prq}{\mbox{PR}_q}
\newc{\tr}{\mbox{tr}}
\newc{\pd}{\partial}
\newc{\qv}{\vec{q}}
\newc{\pv}{\vec{p}}
\newc{\dqv}{\delta\vec{q}}
\newc{\dpv}{\delta\vec{p}}
\newc{\mbq}{\mathbf{q}}
\newc{\mbqp}{\mathbf{q'}}
\newc{\mbpp}{\mathbf{p'}}
\newc{\mbp}{\mathbf{p}}
\newc{\mbn}{\mathbf{\nabla}}
\newc{\dmbq}{\delta \mbq}
\newc{\dmbp}{\delta \mbp}
\newc{\T}{\mathsf{T}}
\newc{\J}{\mathsf{J}}
\newc{\sfL}{\mathsf{L}}
\newc{\C}{\mathsf{C}}
\newc{\B}{\mathsf{M}}
\newc{\V}{\mathsf{V}}

\section{Introduction}

Quantum entanglement is a central property of quantum mechanics that is absent in classical physics.  Studied since Schr\"odinger and the famous paper of Einstein, Podolsky and Rosen (EPR) \cite{epr_paradox}, correlations due to entanglement seem to imply nonlocality. The work of Bell \cite{bell} and others led to inequalities that quantified the extent to which classical correlations can be surpassed. These inequalities were experimentally verified by Aspect {\it et.al}~\cite{aspect}. However, entanglement has been extensively studied more recently as it is a critical resource for quantum computation~\cite{Jozsalinden}, quantum teleportation~\cite{Teleport}, dense coding \cite{Superdense}, and various other quantum information tasks~\cite{Masanes, Pianiwatrous}, and to explain the magnetic properties of some solids \cite{Ghosh}. A well known example of an entangled state is the spin singlet which is a maximally entangled state of two qubits. 

It is known that in a generic or random pure state any of its subsystems is nearly maximally entangled with the complementary system~\cite{Lubkin, Page, Haydenmath, Arul}, where the measure of entanglement is the von Neumann entropy of the subsystem.  Here ``random'' means that the state is sampled uniformly from the unique Haar measure that is invariant under unitary transformations.  Random states occur in many contexts.  For example, they are found as eigenstates of quantum maps whose classical limit is fully chaotic~\cite{HaakeBook}.  For the eigenstates of quantum systems with classically chaotic, continuous Hamiltonian analogs, one must account for an effective dimensionality, i.e. energy window such as the Thouless energy, in addition.  With that proviso, disordered or chaotic ballistic mesoscopic systems~\cite{beenakker97} exhibit randomness in their single particle eigenstates, and in strongly interacting systems such as medium to heavy nuclei with many valence nucleons, there is randomness in the full many-body eigenstates~\cite{brody81}. 
 
\begin{figure}
\begin{center}
 \resizebox{70mm}{!}{\includegraphics{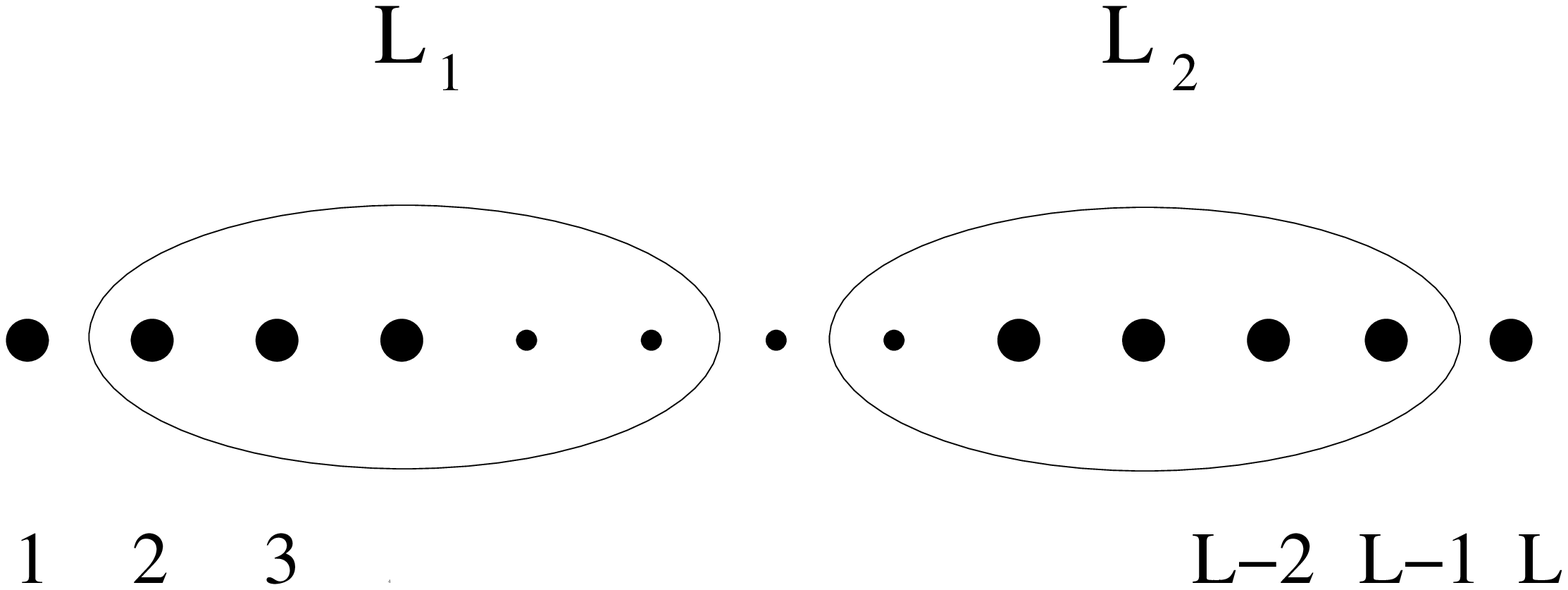}}
   \caption{System of $L$ qubits. $L_1+L_2 < L$}
 \label{fig10} 
\end{center}
 \end{figure}

The interest in this paper is to study the entanglement between two blocks comprising say $L_1$ and $L_2$ qubits in a random pure state $|\psi\rangle$ of $L$ qubits $(L_1+L_2 \leq L)$ (see Fig.~(\ref{fig10})). While one can work with any dimensional Hilbert space, this paper puts the results mostly in terms of collections of ``qubits" or spin-$1/2$ particles, the generalizations being straightforward.
 The reduced density matrix, $\rho_{12}$, of $L_1+L_2$ qubits is obtained by tracing out the remaining $L-L_1-L_2$ qubits:
\begin{equation}
\label{rho12}
 \rho_{12}= \mbox{tr}_{L-L_1-L_2} (|\psi\rangle\langle \psi|).
\end{equation}
The state $\rho_{12}$ is in general a mixed one, {\it i.e.} $\mbox{tr}(\rho_{12})^2 \le\mbox{tr}(\rho_{12})=1$, and the equality holds only if the qubits in the blocks $1$ and $2$ are unentangled from the rest.
A mixed state of a bipartite system is separable if and only if it can be written as 
\begin{equation}
\sum_{i} p_i\; \rho_i^{(1)}\otimes \rho_i^{(2)},\mbox{with}\; p_i \geq 0\; \mbox{and}\; \sum_{i} p_i = 1,
\end{equation}
where $\rho_i^{(1,2)}$ are density matrices of subsystems $1$ and $2$.
Otherwise the state is non-separable, or entangled. Given a general state it is a challenging task
to verify if it is separable or not. 

One simple (partial) test for entanglement is Peres's partial transpose (PT) criterion \cite{Peres}. 
The matrix transpose map $T : \rho \to \rho^T$ is trace preserving and positive, since  for every 
$\rho \geq 0$,  $\rho^{T} \geq 0$.
However its extension $I \otimes T$ to a bipartite system 
(where $I$ is an identity matrix that acts on the first subsystem 
and $T$ acts on the second subsystem) does not preserve positivity. Hence transposition is a positive but not a completely positive map, and can be used to reveal entanglement.
The map $I \otimes T$ is called a partial transposition (PT) since it effects transposition only on the second subsystem keeping the first subsystem unaltered. The test is partial as it leads to necessary but not
sufficient conditions for entanglement.

Given a bipartite system $1$ and $2$ having an orthonormal basis
\{$|i\rangle |\alpha\rangle$\} and density matrix $\rho_{12}$, the PT with respect to the second subsystem, denoted as $\rho_{12}^{T_2}$, is given by the matrix elements:
\begin{equation}
(\rho_{12}^{T_2})_{i\alpha;j\beta} = (\rho_{12})_{i\beta;j\alpha}\;;\;\;\; (\rho_{12})_{i\alpha;j\beta}= 
\langle i|\langle \alpha|\rho_{12}|j\rangle|\beta \rangle.
\end{equation}
Peres's partial transpose (PT) criterion states that if $\rho_{12}^{T_2}$ 
is negative then the state $\rho_{12}$ is entangled. In this case $\rho_{12}$ is said to be a NPT (negative partial transpose) state, 
otherwise it is a PPT (positive partial transpose) state and is guaranteed to be separable only for $2\times2$ and $2\times3$ systems 
\cite{mhorodecki}. Entanglement between $2$ qubits in a mixed state is also given by the concurrence \cite{Wooters,Wootersentform} which takes values from $0$ 
to $1$, where $0$ corresponds to an unentangled or a product state and $1$ corresponds to a maximally entangled state.  For more than two 
qubits or higher dimensional quantum spins in a mixed state, negativity and log-negativity \cite{vidal,logneg} are used as measures of 
entanglement.


In the following, eigenvalues of a density matrix without PT are denoted by $\lambda_i$ and  those of $\rho_{12}^{T_2}$ by $\mu_i$.
Negativity of the state $\rho_{12}$ is defined as the sum of the moduli of the negative eigenvalues of $\rho_{12}^{T_2}$, which is clearly zero for PPT states.
Due to the fact that the trace is preserved under partial transpose the negativity 
is also 
\beq {\cal N}(\rho_{12})= \dfrac{\sum_i |\mu_i| -1}{2}. \label{negativity} \eeq 
Log-negativity is defined as
\begin{equation}\label{logneg}
E_{LN}=\log\big(||\rho_{12}^{T_2}||_1\big)=\log\big(\sum_{i} |\mu_i|\big).
\end{equation} 
If the log-negativity is greater than $0$ then the density matrix is entangled. Otherwise the state $\rho_{12}$ is separable or it could also be bound entangled \cite{mhorodeckibound}. Bound entangled states are entangled but they can not be distilled by means of local operations and classical communication to form a maximally entangled state.
The distribution of the eigenvalues of $\rho_{12}^{T_2}$ is of 
central concern in this paper. Note that the trace of the first and second powers of the density matrix remains unaltered under the PT operation.  The first power to show deviation between the two sets of eigenvalues is the trace of the third power. That is:  
\begin{eqnarray}
\mbox{tr}\left(\rho_{12}\right)= \mbox{tr}\left(\rho_{12}^{T_2}\right)=1,\;\;\nonumber
\mbox{tr}({\rho_{12}}^2) = \mbox{tr}\big[(\rho_{12}^{T_2})^2\big],\;\;
\mbox{tr}({\rho_{12}}^m)\neq \mbox{tr}\big[(\rho_{12}^{T_2})^m\big]\;\;\; m \geq 3.
\label{momentdiffer}
\end{eqnarray}
The average of $\mbox{tr}\big[(\rho_{12}^{T_2})^3\big]$ is explicitly evaluated
further below, for both real and complex states, where the average is over
all the pure states $|\psi\kt$ (see Eq.~(\ref{rho12})) sampled uniformly. Interestingly this quantity is a generalization of one of local unitary invariants studied for three qubits \cite{Kempe99}, and therefore
is of broader interest.

The distribution of the eigenvalues, {\it i.e.}~the density  of $\rho_{12}^{T_2}$ (that of $\mu_i$), is of evident importance in a calculation of the entanglement between 
subsystems $1$ and $2$. In this paper a simple random matrix model is proposed
for the partial transpose, based on the known average of the second
moment. This model quite accurately predicts a transition from dominantly
NPT states to dominantly PPT states as the size of the subspaces $L_1$ and $L_2$ are varied. The transition region is an interesting one wherein the extreme eigenvalues of random matrices determine the nature of the entanglement. Use is made of the 
well-known Tracy-Widom distribution to estimate the fraction of NPT states in the
transition to predominantly PPT ones. The limitations of the simple model are also
pointed out, especially when the skewness of the densities are important and $L_1$ and $L_2$ differ significantly. 

Finally in this paper a dynamical model of three coupled standard maps or rotors is studied, restricting attention to the case when they are classically fully chaotic. 
The eigenstates of such a system are taken to be the pure states in Eq.~(\ref{rho12})
and the entanglement between rotors is studied via the log-negativity measure.
While good agreement is found away from the transition region, there are interesting
deviations in this critical zone. While all standard diagnostics, such as the distribution of the nearest neighbor spacings of the eigenangles, the number variance, the distribution of the eigenvector components, agree with random matrix theory (RMT) to a large extent, deviations are seen with respect to the fraction of NPT states. Stated simply
the dynamical system has systematically more entanglement than predicted by random matrix theory. These tests are perhaps some of the more stringent ones of the Bohigas-Giannoni-Schmit conjecture \cite{Bohigas84} that quantized chaotic systems have spectra whose statistical properties are modeled by those of random matrices. These tests
are stringent as they rely on outliers or extreme eigenvalues. 
 In the large system dimension limit (small effective $\hbar$) there does, however, seem to be a tendency to approach the RMT results.

Works related to the results in this paper have appeared previously. Recently Datta in \cite{Animeshrandom} has studied entanglement of random pure states using negativity \cite{Peres} for equal bipartition which adds to the full system ($L_1=L_2=L$/2) and found that the average negativity is a constant $(0.72037)$ multiple of the maximum possible $(=(2^{L/2}-1)/2)$. This is a reflection of the large entanglement
present in random bipartite pure states, for example as measured by the 
von Neumann entropy of the subsystems \cite{Page}. A calculation presented further below, based on previously derived results in \cite{Nadal11}, gives an 
explicit expression for the average negativity that is also slightly different. In this case the eigenvalues $\mu_i$ of the partial transpose are simply related to the 
eigenvalues of the reduced density matrix and therefore, implicitly, the 
density of the eigenvalues $\mu_i$ had been derived even earlier \cite{Marko}.

If the two subsystems do not make up the full system, Kendon {\it et. al.} in~\cite{Kendonpra,Kendonopt} found numerically that in a typical random pure state the subsystem consisting of $L_1$ and $L_2$ qubits is NPT if $L_1 + L_2 \geq L/2$. Analytically they showed that the lower bound on $L_1+L_2$ for  $\rho_{12}$ to be NPT is $L/3$.  It is shown in this paper that using the simple random matrix model for the partial transpose leads to the bound on $L_1+L_2$ for $\rho_{12}$ to be NPT  is in fact $L/2$.
In \cite{Hilarycircuit} Carteret has given a quantum circuit which can determine the spectrum of $\rho_{12}^{T_2}$ by computing $\mbox{tr}(\rho_{12}^{T_2})^l$ for all $l$'s up to the dimension of $\rho_{12}^{T_2}$. Then from Peres's partial transpose criterion one can determine whether $\rho_{12}$ is NPT or PPT.

The random mixed states studied in this paper are those that arise from a partial trace of random pure states selected according to the Haar measure. Properties of random mixed states generated according to the measure induced by the Bures metric \cite{Sommers04}, have been studied earlier using the von Neumann entropy and purity in \cite{Sommers04,Borot11}. Multipartite entanglement for localized states \cite{Giraud07} \cite{Meyer02}, and multifractal states (using the von Neumann entropy) have also been studied. \cite{Giraud09}. Mathematical work connected to the spectrum of the partial transpose has appeared very recently in the literature~\cite{Aubrun10,Collins11}, which is of a complementary nature, but with some overlap, after much of the present work was done. 

The structure of the paper is as follows. In section ~\ref{knownresult}, some known and relevant results on the reduced density matrix are first summarized.  Rest of this section is a detailed treatment of  the effect of PT on the reduced density matrix, in particular a random matrix model is seen to give rise to the observed Wigner semicircle density of states on PT, and predicts the transition from a predominantly NPT to primarily PPT phase. Further in this same section we calculate the average of the the trace of the third power of the density matrix after PT, and show how it is related to an invariant, the Kempe invariant, that has been studied earlier in the literature. In section~\ref{entanglement} these results are used to find the average log-negativity between two subsystems of the tripartite state.   In subsection~\ref{extstat}, results on extreme value statistics of minima of reduced density matrices after PT are presented, and it is seen how the Tracy-Widom distribution gives rise to the fraction of NPT/PPT states at critical dimensions. In section ~\ref{stdmap}, we compare our results of random states with eigenstates of three coupled quantum standard maps, and find good agreement. 

\section{Statistical properties of the partial transpose}
\label{knownresult}

\subsection{On the reduced density matrix of a subsystem}

If a bipartite quantum system of Hilbert space dimension $N \times M$ ($N\leq M$) is drawn from the ensemble of random pure states then the joint probability density function of the eigenvalues \cite{llyod,Sommers} of the reduced density matrix $\rho_N$ of a subsystem of dimension $N$ is 
\begin{equation}
 P[\{\lambda_i\}]=K_{M,N}\; \delta\bigg(\sum_{i=1}^N \lambda_i-1\bigg)\prod_{i=1}^N \lambda_i^{\frac{\beta}{2}(M-N+1)-1}\prod_{i<j} |\lambda_i-\lambda_j|^{\beta},
\label{jpdfeigen}
\end{equation}
where $\beta$=1, 2 and 4 for real, complex and symplectic case respectively. The normalization constant $K_{M,N}$ is calculated using Selberg's integral \cite{Sommers}. 
 The density of the eigenvalues, for large $N$ and $M$, is given by an appropriately scaled Marcenko-Pastur (MP) function \cite{Marcenko,Nadal11},
\begin{eqnarray}
\begin{split}
f(\lambda)& =  \frac{NQ}{2\pi} \frac{\sqrt{(\lambda_{+} -\lambda )(\lambda- \lambda_{-} )}}{\lambda}\\
\lambda_{\pm} &= \frac{1}{N}\bigg(1+\frac{1}{Q} \pm \frac{2}{\sqrt{Q}}\bigg), 
\end{split}
\label{MPfunct1}
\end{eqnarray}
where $\lambda\in [\lambda_{-}, \lambda_{+} ]$, $Q=M/N$ and $Nf(\lambda)d\lambda$ is the number of eigenvalues in the range 
$\lambda$ to $\lambda+d\lambda$. 
For $Q=1$ there is a divergence at the origin. For $Q>1$ the eigenvalues are bounded away 
from zero.  

The purity of the subsystem density matrix $\rho_N$ is always larger than $1/N$ and less than $1$. The minimum value being attained when $\rho_N$ is maximally mixed, and the maximum when the subsystems are unentangled. 
The average purity of the subsystem $\rho_N$ is given by \cite{Lubkin} 
\begin{equation}
\left \langle \mbox{tr}\big[(\rho_N)^2\big] \right \rangle= \frac{N+M}{NM+1}\approx  \frac{1}{N}+\frac{1}{M},
\label{avepurity}
\end{equation}
the last approximation being valid for $N,M \gg 1$.
The subsystem entropy is a good measure of bipartite pure state entanglement and remarkably there is an exact formula for its average evaluated over the probability density in  Eq.~(\ref{jpdfeigen}) \cite{Page,Sen,Jorge}. 
\begin{equation}
\left \langle-\mbox{tr}(\rho_N\log\rho_N)\right\rangle\, =\, 
 \sum_{m=M+1}^{NM} \frac{1}{m}-\frac{N-1}{2M} \approx \log(N)-\frac{N}{2M}\;\; \mbox{for}\;\; 1 \ll N \leq M.
\end{equation}
In terms of an interpretation, there is practically very little information about the full pure state in a subsystem, to be more precise there is less than one-half unit of information on average in the smaller subsystem of a total system in a random pure state. The maximum entanglement being $\log(N)$, there is near maximal entanglement between any two subsystems of a random state.

\subsection{Effect of PT On Reduced Density Matrix} 
\label{effectPT}

\begin{figure}
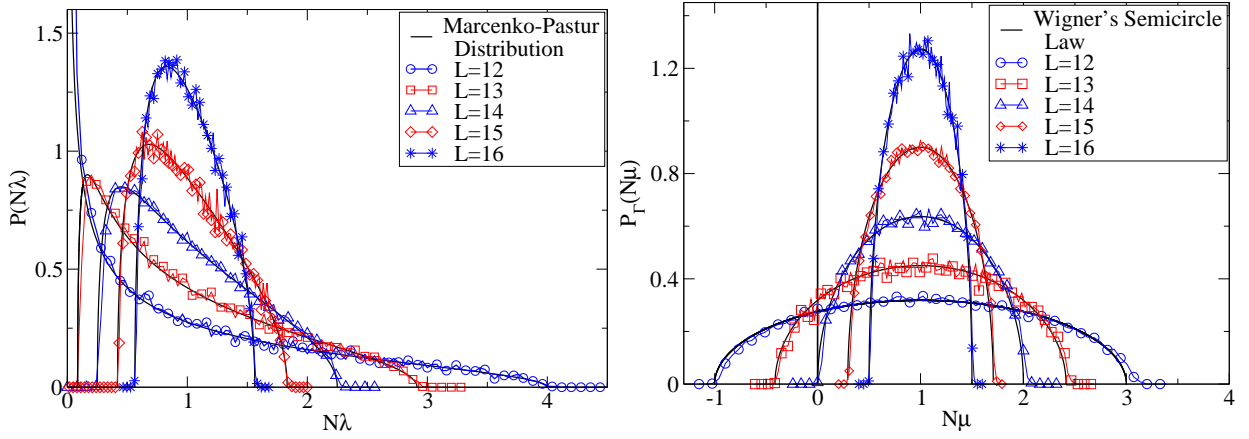

\resizebox{80mm}{!}{\includegraphics{wishart.eps}}
\resizebox{82mm}{!}{\includegraphics{wignerdatafit2.eps}}
\caption{(Color online) Density of states of $\rho_{12}$ (left) and $\rho_{12}^{T_2}$ (right) for $L_1=L_2=3$, and where $L$ is the total number of qubits. 
A vertical line at the origin has been shown in the right figure to draw attention to the negative part of the spectrum. In each case $250$ complex random states are used. }
\label{MPandWigner} 
\end{figure}

Reverting back to the notation of $\rho_{12}$ as the reduced density matrix, while its density 
of eigenvalues is the scaled Marcenko-Pastur distribution in Eq.~(\ref{MPfunct1}), we are 
 interested in the spectrum of its partial transpose, $\rho_{12}^{T_2}$. 
It is numerically found that for $L_1=L_2$, the eigenvalue density of $\rho_{12}^{T_2}$ 
fits the well-known Wigner's semicircle law for any $L$ such that $L_1+L_2 \ll L$; see Figs.~(\ref{MPandWigner},\ref{fig1}). Oscillations are found about the semicircle for very small values of $L_i$, just as 
in the case of the canonical ensembles of RMT \cite{Mehtabook,Forresterbook}.
The figure shows results for $L_1=L_2=3$ and varying $L$ from $12$ to $16$. The semicircle's are
fit according to a center (or shift) and width that is discussed further below. The rather good agreement with the semicircle law for the spectrum of the partial transpose is evident. 
For instance, in the case when $L_1=L_2=L/4$ (corresponding to the case $L=12$ in Fig.~(\ref{MPandWigner})) the rescaled eigenvalues $x=N\mu$ fit the following formula:
\begin{equation}
\label{WignerLby4}
P_{\Gamma}(x)=\frac{1}{2\pi}\sqrt{4-(x-1)^2} \;\; \mbox{where} \; -1\le x\le3 \;\;\mbox{and} \;N=2^{L_1+L_2}.
\end{equation}

Recently Aubrun \cite{Aubrun10} has used the binary correlation method \cite{brody81} to find a shifted Wigner's semicircle law under PT. We however use an approximate and simple model that enables us to see the transition that is observed in Fig.~(\ref{MPandWigner})) when the total number of qubits is $L=14$. The said transition is from a predominantly PPT phase when $L_1+L_2 < L/2-1$ to a predominantly NPT one when $L_1+L_2 > L/2-1$.
The critical case is an interesting one that is fit for the application of extreme value statistics to find the fraction of NPT states. In this case the semicircle lower bound is at $0$. When $L$ is odd however one finds that there is no $L_1+L_2$ which is critical in this sense; instead for $L_1+L_2 \le (L-3)/2$ the states
are predominantly PPT and if  $L_1+L_2 > (L-3)/2$ are predominantly NPT. If one is given a certain number of qubits $L_1+L_2$, then there is always the case when the total number of qubits is $L=2(L_1+L_2)+2$ which is critical. In this work most of the calculations are for $L$ even, and there is a critical subspace dimension $L_1+L_2$=$L/2-1$. When $L_1+L_2=L$, so that the ``subsystem" $1+2$ is, in fact, the whole
system and is in a pure state, much can be said about the spectrum of the partial transpose. This case, discussed later in this paper, has a density of states that is {\it not} the Wigner semicircle.  However, a semicircle is obtained even from small deviations of $L_1+L_2$ away from $L$.

\subsubsection{Degree of partial derangement in the partial transpose}
 
 The PT operation partially rearranges the positive matrix $\rho_{12}$ through selective exchange of matrix elements. One may expect that the extent of such a rearrangement will be connected with a deviation from the Marcenko-Pastur distribution and approach toward the semicircle law.  In other words, the number of elements exchanged by the PT operation results in a loss of the particular correlation among matrix elements necessary to make the original matrix positive.  However, the rearrangements do preserve the Hermitian nature of the matrices.   Additionally, for a density matrix of $M$ qubits, the eigenvalues of the matrix obtained after PT on $k$ qubits are the same as after doing the partial transpose on the complementary $M-k$ qubits. Thus, the range $0 \le k \le M/2$ is the full range of interest. 

\begin{figure}
\begin{center}
\resizebox{90mm}{!}{\includegraphics{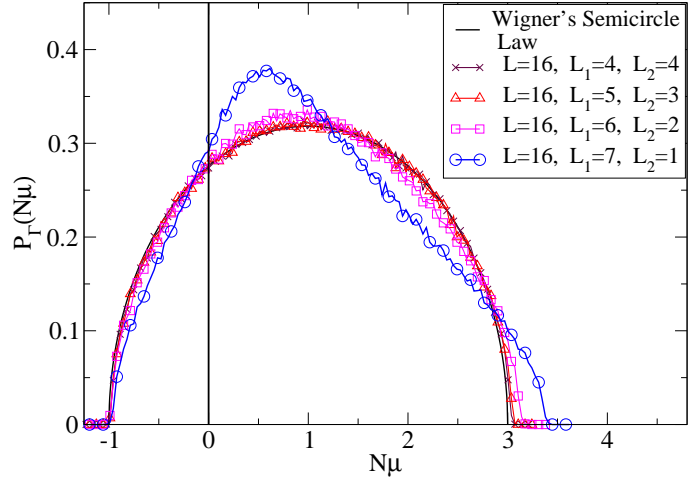}} 
\end{center}
\begin{center}
\caption{(Color online) Density of states after partial transpose for various $L_1$, $L_2$ and fixed $L$. The skewness is minimum for $L_1=L_2=4$  and maximum when $L_1=1$ and $L_2=7$. Except for the case when $L_1=1$ and $L_2=7$ all other cases are close to Wigner's 
semicircle law.} 
\label{skewness} 
\end{center}
 \end{figure}

First divide the whole matrix of dimension $2^M \times 2^M$ into matrices  
of dimension $2^k \times 2^k$; the number of such matrices being $2^{2(M-k)}$. PT on $k$ qubits is a full transpose on these $2^k \times 2^k$ matrices.
 Therefore the number of elements getting exchanged after PT is 
 \begin{equation}
\#=2^{2M}-2^{2(M-k)} 2^k=2^{2M}(1-2^{-k}).
\end{equation}
This number, which is evidently the same for whether the density matrix is real or complex, is maximum when $k=M/2$, and therefore one can expect the maximum loss of correlation among matrix elements
of $\rho_{12}$ and the development of the Wigner semicircle law.  When $k$ is smaller one still 
obtains qualitatively different spectra depending on the density matrix. In Fig.~(\ref{skewness}) this is  
seen, with $M=L_1+L_2=8$ and $k=L_2$. As the number of qubits $k$ varies from $1$ to $4$. The obtained densities on PT are all very similar except for the extreme case of $k=L_2=1$, when the skewness is more apparent. Remarkably the minimum of the distributions remain unchanged even as the
maximum shifts slightly. The question of the skewness is addressed further below.
 
 \subsubsection{A model for the shifted semicircles}
 
 
A simple model for the spectral density of $\rho_{12}^{T_2}$, the PT of a density matrix, is suggested by the fact that the first two moments do not change under the operation of PT.  As a semicircular density depends on just the two moments of mean and variance, it is proposed to shift and scale the semicircle of the Gaussian random ensembles to match the first two moments of $\rho_{12}$ ( or equivalently $\rho_{12}^{T_2}$).  In particular, we assume that these random matrices belong to the  Gaussian Unitary Ensemble (GUE). Thus consider
\beq
B=A+\dfrac{I_N}{N}
\label{model}
\eeq
where $A$ is a $N\times N$ GUE random matrix with the necessary matrix element variance to match the variance of $\rho_{12}$ and $I_N$ is the identity matrix. It follows that  $\langle \mbox{tr}(B)\rangle =1$ since $\langle \mbox{tr}(A)\rangle =0$, where the angular brackets indicate the ensemble average.  The fact that $\mbox{tr}(B)$ is not exactly equal to unity for each and every member of the ensemble would be expected to have an influence only in the case of very small dimensional cases.

As the eigenvalues of $B$ are all those of $A$ shifted by $1/N$, it is sufficient to consider the spectrum of $A$. Under the assumption that it is from the GUE it follows that the density of eigenvalues of $B$ is (for large $N$)
\beq
 P(\mu)=\frac{2}{\pi R^2}\sqrt{R^2-\bigg(\mu-\frac{1}{N}\bigg)^2},\; -R+\frac{1}{N} < \mu < R+\frac{1}{N},
\label{wignerdist}
\eeq
where 
\begin{equation}
R=2\sqrt{\dfrac{1}{N} \langle\mbox{tr}(A^2)\rangle}= 2 \sqrt{\frac{1}{N}\br \tr(\rho_{12}^2) \kt \, -\, \frac{1}{N^2}}.
\end{equation} 
Use is now made of the approximate form of the average purity in Eq.~(\ref{avepurity}) to derive that
\beq
\label{R}
 \langle\mbox{tr}(A^2)\rangle = \frac{1}{N_3},\;
R=\dfrac{2}{\sqrt{N_3\, N}}= 2^{-L/2+1},
\eeq
where one has also used that $N_3=2^{L_3}$, $N=2^{L_1+L_2}$ and $L_1+L_2+L_3=L$.
If the scaled variable $x= \mu N$ is used, the resultant semicircular probability density has a shift of $1$
and a rescaled ``radius" $\tilde{R} = N R= 2^{L_1+L_2-(L/2-1)}$. Explicitly:
\beq
P_{\Gamma}(x)=\frac{2}{\pi \tilde{R}^2}\sqrt{\tilde{R}^2-(x-1)^2},\; \;1-\tilde{R} < x < 1+\tilde{R}.
\label{wignerdistscaled}
\eeq
This is the Wigner semicircle law that has been used in Figs.~(\ref{MPandWigner}, \ref{skewness}) and
illustrates how well this simple model works.

Moreover this treatment gives the PPT~-~NPT transition as well. For if $L$ is an even integer and
$L_1+L_2=L/2-1$ then $\tilde{R}=1$ and the radius of the (rescaled) semicircle is such that the lower limit
is exactly at $0$. For any $L_1+L_2> L/2-1$ the radius is larger than unity and there are NPT states,
while in the opposite case the lower bound is such that there are predominantly PPT states. Thus the transition is clearly indicated in the model of the partial transpose as a shifted random matrix of the GUE kind. If $L$ is odd, it is clear that there are no $L_1$, $L_2$ such that the radius is unity, but it is
easy to find that when $L_1+L_2=(L-1)/2$, the radius is $\sqrt{2}$ and hence the states are predominantly
NPT, while when $L_1+L_2=(L-3)/2$, the radius is $1/\sqrt{2}$ and hence the states are predominantly
PPT. These are indeed statements that are based on the model introduced above, but are well corroborated by numerical simulations as presented for example in Table \ref{table0029}.

An additional interesting feature is that the model predicts that the range of the eigenvalues
is the same both before and after the PT. Namely
\beq
N(\lambda_{+} - \lambda_{-}) = 2 \, \tilde{R} = 4 \sqrt{N/N_3}= 2^{L_1+L_2-(L/2-2)},
\eeq
where $\lambda_{\pm}$ are the limits of the Marcenko-Pastur distribution in Eq.~(\ref{MPfunct1}). This is borne out in Fig.~(\ref{MPandWigner}). While this is not an exact equality, it seems to be nearly true statistically. Extreme deviations from this will occur when the subsystem $1+2$ is nearly pure or pure,
a case we will discuss later.
Is there some characteristic of the density matrix $\rho_{12}$ that signals the PPT~-~NPT transition?
Note that when $L_1+L_2=L/2$, $\rho_{12}$ has a density of states that diverges at $0$, see Fig.~(\ref{MPandWigner}), and for $L_1+L_2>L/2$, the density matrix is rank deficient.Whereas the critical case as far as this transition goes is at $L_1+L_2=L/2-1$ when the density of states of $\rho_{12}$ does not diverge at zero.  While the rank of the density matrix mattered in the case
of an entanglement transition observed for definite-particle states recently \cite{Vikram11}, it seems
to be not exactly the case here, as there is a case when the density of states of $\rho_{12}$ is bounded
away from zero, but its partial transpose has a significant measure of negative eigenvalues and is predominantly NPT.

As is apparent from the Fig.~(\ref{skewness}) the semicircle is not obtained when one
of the subspaces is of very low dimensions, although interestingly even in this case the 
minimum eigenvalues after PT remains nearly the same. We limit most of our discussions
to those cases where the semicircle law is approximately valid. Another instance
where the semicircle law is not valid is when the third subspace has no qubits, that is 
the state $\rho_{12}$ is itself pure. This case will be discussed in the next section. More work
needs to do be done in elucidating the boundaries of the applicability of various densities after PT.
For the sake of clarity averages calculated using the shifted GUE model are denoted as $\br \cdots \kt _M$,
while averages calculated over the ensemble of random pure states is simply
$\br \cdots  \kt$. 

\subsubsection{The third moment, the Kempe invariant, and the skewness}
 
The lowest ordered moment which changes after PT is the third moment {\it i.e.} 
$\mbox{tr}\big[(\rho_{12})^3\big]$ $\neq$ $\mbox{tr} \big[(\rho_{12}^{T_2})^3\big]$ and it is therefore interesting to calculate the {\it exact} ensemble average $\br \mbox{tr}\big[(\rho_{12}^{T_2})^3\big] \kt$, and compare it with that of the simple model above.
In the case of complex random pure states of $L=L_1+L_2+L_3$ qubits, we find 
\begin{equation}
\langle\mbox{tr}\big(\rho_{12}^{T_2}\big)^3\rangle =\dfrac{N_1^2+N_2^2+N_3^2+3N_1N_2N_3}{
(N_1N_2N_3+1)(N_1N_2N_3+2)}=\frac{N_1^2+N_2^2+2^{2(L-L_1-L_2)}+3\times 2^L} {(2^L+1)(2^L+2)},
\label{tr3cmplx}
\end{equation}
where $N_i=2^{L_i}$. Details of the derivation are relegated to the Appendix. In contrast, prior to PT,
\begin{equation}
\langle\mbox{tr}\big(\rho_{12}\big)^3\rangle =\frac{ N_1^2N_2^2+N_3^2+3\;N_1 N_2N_3+1}{(N_1 N_2N_3+1)(N_1N_2 N_3 +2)},
\label{tr3rho}
\end{equation}
so that
\begin{equation}
\langle \mbox{tr}\big(\rho_{12}\big)^3 \,-\, \mbox{tr}\big(\rho_{12}^{T_2}\big)^3\rangle =\frac{ (N_1^2-1)(N_2^2-1)}{(N_1 N_2N_3+1)(N_1N_2 N_3 +2)}.
\end{equation}
Thus, on average the third moment after PT is smaller than that before. The equation in Eq.~(\ref{tr3rho}) is a special case of Eq.~(\ref{tr3cmplx}), with the identification of $N_2 \equiv 1$ and $N_1 \equiv N_1N_2$, as the original density matrix is the same as a partial transpose over zero qubits. 

The ensemble average of the third moment after PT has a permutation symmetry as is clear from Eq.~(\ref{tr3cmplx}). Quite remarkably,
this is true for {\it every} realization in the ensemble, and is a 
property therefore of pure states split in a tripartite way. To be explicit, in this case the following can be shown to be true:
 \beq
 \label{kempe}
 \mbox{tr}\left(\rho_{12}^{T_2}\right)^3 =
\mbox{tr}\left(\rho_{23}^{T_3}\right)^3 = 
\mbox{tr}\left(\rho_{31}^{T_1}\right)^3 
\eeq
Note that there is no such constraint for the density matrices $\rho_{12}$, $\rho_{23}$,
and $\rho_{13}$ themselves. 
To our knowledge, this has been identified as one local 
unitary invariant for the case of three qubits \cite{Hilarycircuit}, but not for general tripartite systems. In the case of three qubits this quantity, which has however been written differently, has been called the ``Kempe invariant" and denoted as $I_5$ \cite{Kempe99,Sudbery01,Williamson11}.

For completeness a proof is now supplied for the identity in Eq.~(\ref{kempe}). Let the pure tripartite state and its adjoint be written in a standard basis as
\beq
|\psi\kt =\sum_{jkl}\psi_{jkl}|jkl\kt, \;\; \br \psi| =\sum_{jkl}\overline{\psi}^{jkl} \br jkl|,
\eeq
where $1\le j \le N_1$, $1\le k \le N_2$, $1\le l \le N_3$, and $\overline{\psi}$ is the complex conjugate of $\psi$. The following then ensues (repeated indices are summed over):
\beq
\label{wavefunction1}
\rho_{12} = \sum_{jk,j'k'} \psi_{jkl} \overline{\psi}^{j'k'l} |jk\kt \br j'k'|, \; 
\rho_{12}^{T_2} = \sum_{jk,j'k'} \psi_{jk'l} \overline{\psi}^{j'kl} |jk\kt \br j'k'|.
\eeq
\beq
\label{tr12ptcube}
 \mbox{tr}\left(\rho_{12}^{T_2}\right)^3=\psi_{jk'l}\overline{\psi}^{j'kl} \psi_{j'k''l'}\overline{\psi}^{j''k'l'} \psi_{j''kl''}\overline{\psi}^{jk''l''}.
\eeq
Similarly it follows on tracing out the second system and taking the partial transpose with the first that:
\beq
\label{tr31ptcube}
\mbox{tr}\left(\rho_{31}^{T_1}\right)^3=\psi_{j'kl}\overline{\psi}^{jkl'} \psi_{j''k'l'}\overline{\psi}^{j'k'l''} \psi_{jk''l''}\overline{\psi}^{j''k''l}.
\eeq
To see the equality of Eq.~(\ref{tr12ptcube}) and Eq.~(\ref{tr31ptcube}) the following permutation of the dummy indices suffices: $(j \rarrow j'', \, j'' \rarrow j',\, j' \rarrow j)$, and $(k \rarrow k',\, k' \rarrow k'',\, k'' \rarrow k)$. It seems somewhat unusual to express the Kempe invariant in terms of the partial transpose, but this indeed seems to be a simple way of doing so. 

Since the quantity is both a local unitary invariant and invariant under permutation of the systems it serves as some kind
of entanglement measure in itself. Thus the average of this quantity as found in Eq.~(\ref{tr3cmplx}) is of larger interest as well. The average Kempe invariant of three qubits is $2/5$ while for three qutrits it is $27/203$.
According to \cite{Sudbery01} it is a measure of bipartite entanglement, indeed it is possible that it is some overall measure of entanglement between any 
pair of the tripartite system. For the generalized W-state:
\beq
\label{Wstate}
|\psi_W\kt = \alpha |001\kt + \beta |010\kt +\gamma |100\kt
\eeq
this invariant is 
\beq
\label{Winvariant}
\mbox{tr}\left(\rho_{12}^{T_2}\right)^3
=\alpha^6+\beta^6+\gamma^6+3\alpha^2\beta^2\gamma^2,
\eeq
which clearly displays the permutation symmetry on interchange of qubits.
It follows that $2/9 \le \mbox{tr}\left(\rho_{12}^{T_2}\right)^3  \le 1$, the smallest value of the invariant corresponding to the W-state with $\alpha=\beta=\gamma=1/\sqrt{3} $. A special case of interest is when say 
$\alpha=0$, but $\beta, \gamma \ne 0$, when only the first two qubits are entangled with each other. 
It is not hard to show (see Appendix \ref{highmoments}) that in this case all the 
odd moments $\mbox{tr}\left(\rho_{12}^{T_2}\right)^{(2k+1)}$, $k=0,1,2,\ldots$
are permutation symmetric, although the third qubit is clearly special. This property of the higher moments being permutation symmetric is lost when all three qubits are entangled. 

Although it can be shown that in general  $\mbox{tr}\left(\rho_{12}^{T_2}\right)^n = \mbox{tr}\left(\rho_{13}^{T_3}\right)^n$ only for $n=1, 3$, (Appendix \ref{highmoments}), one may also simply offer an example as provided by the W-state with $\alpha=\sqrt{3/7}$, $\beta=\gamma=\sqrt{2/7}$. 
This leads to (see Appendix \ref{app:counterexample} for details)
\begin{eqnarray}
\label{nthmoment}
\begin{split}
\mbox{tr}\left(\rho_{12}^{T_2}\right)^n &= (2/7)^n+(2/7)^n+(4/7)^n+(-1/7)^n\\ 
\mbox{tr}\left(\rho_{13}^{T_3}\right)^n &= \left(3/7\right)^n+\left(2/7\right)^n+\left(\left(1+\sqrt{7}\right)/7\right)^n
+\left(\left(1-\sqrt{7}\right)/7\right)^n
\end{split}
\end{eqnarray}
 and to two integer sequences whose $n^{th}$ terms are $t_n$ and $t_n'$.
These sequences are important as $\mbox{tr}\left(\rho_{13}^{T_3}\right)^n \ne \mbox{tr}\left(\rho_{12}^{T_2}\right)^n$ iff 
$t_n \ne t_n'$. The $n^{th}$ term of these sequences are
\beq
t_n=3^n +(1-\sqrt{7})^n +(1+\sqrt{7})^n, \;\;t_n' =2^n +4^n +(-1)^n,
\eeq
which generate the sets $\{5, 25, 71,  265, 875, 3097,\cdots\}$ and 
$\{5, 21, 71,  273, 1055, 4161,\cdots\}$ respectively.
The fact that the trace and the third moment are permutation symmetric is
reflected in the equalities $t_1=t_1'$ and $t_3=t_3'$. It can be shown (see Appendix \ref{app:counterexample}) that indeed $t_n \neq t_n'$ for any other values of $n$ and hence $\mbox{tr}\left(\rho_{12}^{T_2}\right)^n = \mbox{tr}\left(\rho_{13}^{T_3}\right)^n$ iff $n=1$ or $n=3$.

Ending this digression into the Kempe invariant {\it per se}, one may also compare its average with the third moment for the shifted GUE matrices in the model of Eq.~(\ref{model}).  The third moment is approximately:
\beq
\br \mbox{tr}(B^3) \kt_M \approx \dfrac{3}{N_1N_2N_3} + \dfrac{1}{N_1^2N_2^2},
\eeq 
neglecting higher order terms.
While this does have the correct leading behavior, it is not the same as the exact moment, and does
not also possess the permutation symmetry noted above. The difference between this result and the exact moment is of a lower order than the exact moment:
\beq
\langle \mbox{tr}\big(\rho_{12}^{T_2}\big)^3\kt  \, -\,\br\mbox{tr}(B^3)\rangle_M \, \approx \, \dfrac{1}{N_3^2} \left( \dfrac{1}{N_1^2}+ \dfrac{1}{N_2^2}\right).
\eeq
The skewness of the density of states is zero for the shifted GUE ensemble, but is nonzero for the PT of the density matrices. 
Skewness $\gamma$ of a distribution is the normalized third central moment:
\begin{equation}
\gamma= \dfrac{1}{N} \sum_{i=1}^N \left(\dfrac{\mu_i -\overline{\mu}}{\sigma}\right)^3,
\label{skewformana}
\end{equation}
 where $N$ is number of elements and $\sigma$ is standard deviation of
the sample. Using $\overline{\mu}= \mbox{tr}(\rho_{12})/N=1/N$, and 
$\sigma^2 =  \mbox{tr}(\rho_{12}^2)/N-\overline{\mu}^2 \approx 1/(N_1N_2N_3)$ (recall
that $N \equiv N_1N_2$) leads to 
\begin{equation}
\gamma \approx \dfrac{1}{\sqrt{N_1N_2N_3}} \left( \frac{N_2}{N_1}+ \frac{N_1}{N_2}\right).
\label{skewform}
\end{equation}

In terms of number of qubits the result is that for large $L_1$, $L_2$ and $L$
\beq
\label{skewform}
\gamma= \left\{
\begin{array}{ll}
2^{-L/2}(2^{L_1-L_2}+2^{L_2-L_1}) & \mbox{(complex states)}\\
2^{-L/2}[2^{L_1-L_2}+2^{L_2-L_1}+3(2^{-L_1}+2^{-L_2})] &\mbox{(real states).}
\end{array} \right.
\eeq
The case of real states is stated {\it only} for completeness, but all the results presented 
are for the complex case. The difference in the real case is also dealt with in the Appendix.
Thus it follows that for a given $L$ and $L_1+L_2$, the skewness is a minimum 
for the symmetric case $L_1=L_2$ (refer Fig.~(\ref{skewness})) when it equals $2^{-L/2+1}$  and tends to zero as 
$L \rightarrow \infty$. In terms of the system dimensions it is also clear that
when $N_1/N_2$ is fixed and the system dimensions tend to infinity the skewness tends to zero. To give some numbers, 
for the cases shown in Fig.~(\ref{skewness}) with $L=16$ and $L_1+L_2=8$, the average skewness $\gamma= .0078,\, .0165, \, .0628,$ and $.2509$ when $L_1=4,3,2$ and $1$ respectively. These match well with the analytical estimates above,
given that the numerical values were from 1000 trials with complex states.
 

 \section{Entanglement} 
\label{entanglement}

With the statistical properties of the partial transpose, entanglement between the subspaces
1 and 2 can now be calculated via the negativity or the logarithmic negativity.
 
\subsection{Logarithmic negativity}
The average log-negativity between two subsystems 1 and 2 of dimensions $N_1$ and $N_2$
is now sought. It is assumed that the system 1+2 is a subsystem of a random pure state
in a $N_1N_2N_3$ dimensional Hilbert space, and the average is over the ensemble of 
uniformly distributed pure states in this space.
Recall that the log-negativity is given by 
$E_{LN}=\log\big(\sum_i |\mu_i|\big)= \log\big(1-2 \sum_{i; \mu_i<0} \mu_i \big) =\log\big(1-2\int_{\mu<0} \mu \;P(\mu)\; d\mu\big)$
where $P(\mu)$ is Wigner's semicircle given in Eq.~(\ref{wignerdist}). Thus

\begin{figure}
\begin{center}
       \resizebox{85mm}{!}{\includegraphics{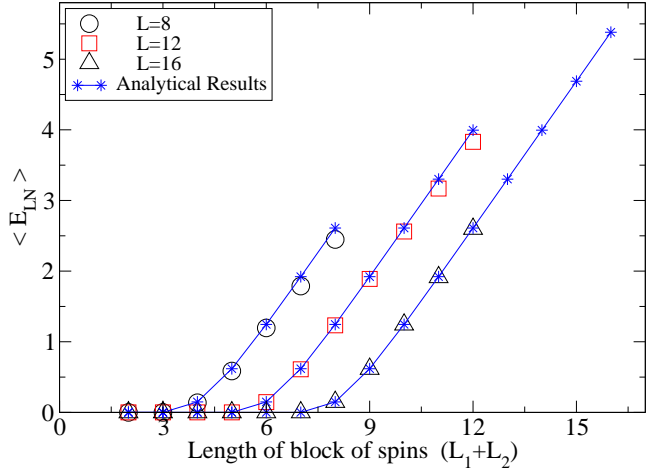}} 
   \caption{(Color online) The average entanglement in states sampled according to the Haar measure, as measured by the log-negativity, between blocks of various sizes compared with the analytical result based on the model ($\br E_{LN}\kt_M$ in Eq.~(\ref{lognegform})). The sizes $L_1$ and $L_2$ are such that if $L_1+L_2$ is even, they are equal and if it is odd, they differ by 1.
   The average log-negativity is zero when $L_1+L_2<L/2-1$.} 
\label{lVsL} 
\end{center}
 \end{figure}

\beq
\br E_{LN}\kt_M =   \log \Bigg[\frac{2}{\pi} \sin^{-1}\Big(\frac{1}{\tilde{R}}\Big)+
      \frac{2}{3 \pi \tilde{R}}\sqrt{1-\frac{1}{\tilde{R}^2}} \left(1+2 \tilde{R}^2\right) \Bigg], 
\label{lognegform}
\eeq
where, as defined earlier, $\tilde{R} = NR= 2\sqrt{N_1N_2/N_3}$. This is valid for 
$\tilde{R}>1$. When $\tilde{R}=1$ (or $N_3=4 N_1 N_2$), which is the critical case, this formula gives zero for the average log-negativity, while this is not true as discussed below. When $\tilde{R}<1$, PPT states are predominantly obtained and $\br E_{LN}\kt =0$. 
Fig.~(\ref{lVsL}) shows how well Eq.~(\ref{lognegform}) works.

On the other hand for $\tilde{R} \gg1$, deep in the NPT regime, Eq.~(\ref{lognegform}) gives
\beq
\br E_{LN} \kt_M \approx \log\left( \f{8}{3 \pi} \sqrt{\df{N_1N_2}{N_3}}\right).
\label{avglogneg}
\eeq
One may compare this with the maximum possible log-negativity of a state in Hilbert space of dimension $N_1N_2$, as well as the average log-negativity over pure states of subsystem 1+2.
From Fig.~(\ref{lVsL}) it is clear that there are deviations when $L_1+L_2=L$, that is the 
subsystem 1+2 is pure. If we put $N_3=1$ (equivalently, $L_3=0$ or $L_1+L_2=L$) we get 
that $\br E_{LN} \kt_M \approx \log(8 \sqrt{N_1N_2}/(3 \pi))$. Now we present a more accurate and 
independent derivation of the average log-negativity in this case.

\subsubsection*{Entanglement when $\rho_{12}$ is pure} 

Bipartite entanglement in a random pure state is known to be very large. 
When $L_3=0$ the $L_1+L_2$ qubits are in a pure state. The eigenvalues of $\rho_{12}^{T_2}$
are directly related to the eigenvalues of $\rho_1$, the reduced density matrix of subsystem 1.
If the eigenvalues of the latter are $\lambda_i$, $i=1, \cdots, 2^{L_1}$, from Schmidt decomposition we have that ($\rho_{12}=|\psi_{12}\kt \br \psi_{12}|$): 
\beq
\label{eq:PTeigenvaluesPurestate}
|\psi_{12}\kt = \sum_i\sqrt{\lambda_i} |\phi_i^{(1)}\kt |\phi_i^{(2)}\kt, \; \rho_{12}^{T_2}=\sum_{ij} \sqrt{\lambda_i \lambda_j } |\phi_i^{(1)} \kt |\phi_j^{(2)}\kt  \br \phi_j ^{(1)}| \br \phi_i^{(2)}|. 
\eeq
It follows that the eigenvalues of $\rho_{12}^{T_2}$ are $\{ \ld_i,\,\pm \sqrt{\ld_i  \ld_j}; \,  i\ne j,\, i,j=1,\cdots,2^{L_1} \}$, the eigenvectors being $|\phi_i^{(1)}\kt |\phi_i^{(2)}\kt$ and 
$|\phi_i^{(1)}\kt |\phi_j^{(2)}\kt\pm |\phi_j^{(1)}\kt |\phi_i^{(2)}\kt$ when $i\ne j$. The rest of
the eigenvalues, if any, are zero.

\begin{figure}
\resizebox{82mm}{!}{\includegraphics{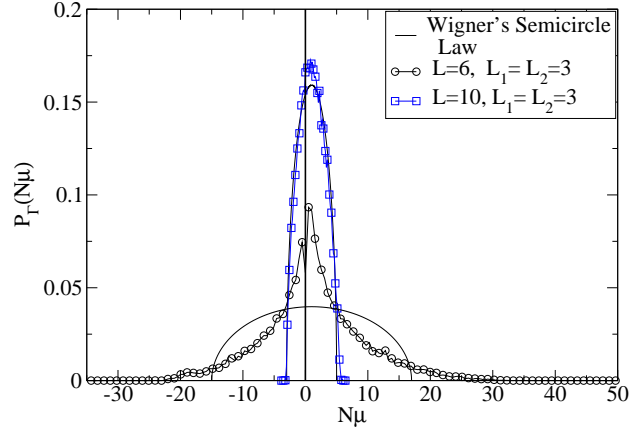}}
\caption{(Color online) Density of (scaled) $\mu$, the eigenvalues after PT for two different cases.
When $L=10$ and $L_1=L_2=3$ ,Wigner's semicircle law is a good fit. In the case when $L=6$ and $L_1=L_2=3$ a different distribution is obtained. A corresponding semicircle of radius $16$ is also shown. In general a semicircle is obtained for $L_1+L_2 \ll L$.}
\label{fig1} 
\end{figure}

Thus the average log-negativity  is found as 
\beq
\br E_{LN} \kt = \left \br \log \left( \sum_{i=1}^{N_1N_2} |\mu_i| \right) \right \kt =  \left \br\log \left( \sum_{i=1}^{N_1} \sqrt{\ld_i} \right)^2\right \kt \approx \log( \kappa^2 N_1),
\eeq
where $N_2 \ge N_1$ and the last approximation is valid for $N_1\gg1$.
Here the number $\kappa$ is found on using the Marcenko-Pastur distribution of Eq.~(\ref{MPfunct1}) to be
\beq
\begin{split}
\kappa=& \f{Q}{2 \pi} \int_{x_{-}}^{x_{+}} \sqrt{\dfrac{(x_{+}-x)(x-x_{-})}{x}} dx = \\ &(\sqrt{Q}-1) \left \{
{_2F_1}\left[\f{1}{2},-\f{1}{2},2,\f{-4\sqrt{Q}}{(\sqrt{Q}-1)^2}\right] -  {_2F_1} \left[\f{1}{2},\f{1}{2},2,\f{-4\sqrt{Q}}{(\sqrt{Q}-1)^2} \right] \right\},
\end{split}
\label{kappa}
\eeq
where $x_{\pm}= (1 \pm 1/\sqrt{Q})^2$, and $Q=N_2/N_1>1$.
In the special case when $N_1=N_2$, or $Q=1$, the integral in Eq.~(\ref{kappa}) is elementary and leads to 
\beq
\br E_{LN} \kt  \approx \log\left[ \left(\frac{8}{3 \pi} \right)^2 N_1 \right].
\eeq
This can be compared with Eq.~(\ref{avglogneg}) which comes from a semicircle and the simple model. One sees that they are indeed close, but intriguingly differ by a square in the constant.
We do not expect the semicircle to hold in the case when $N_3=1$ or $L_1+L_2=L$. Indeed
the density of the partial transposed spectrum is known in this case \cite{Marko}.
Fig.~(\ref{fig1}) shows the deviation of the spectrum from the semicircle when the system
$1+2$ is a pure state, and a distinctive cusp distribution is seen. The same figure also shows how poorly a semicircle with the same first two moments will fare. In the case when $L=6$
and $L_1=L_2=3$ the scaled radius of a purported semicircle will be (from Eq.~(\ref{wignerdistscaled})) $\tilde{R}=2^{(L/2+1)}=16$, which is also shown for comparison.
A more detailed study of the transition from the cusp to the semicircle is warranted, but not carried forward here.

We also note parenthetically that the average negativity, as defined in Eq.~(\ref{negativity}) when $\rho_{12}$ is pure and $N_1=N_2$ is given by
\beq
\br N(\rho_{12})\kt = \left \br \df{\sum_{i}|\mu_i|-1}{2} \right \kt \approx \f{1}{2}\left[ \left(\f{8}{3 \pi}\right)^2 N_1-1\right].
\label{avgnegativity}
\eeq
This maybe compared with an equation for the same quantity in \cite{Animeshrandom} which states that $ \br N(\rho_{12})\kt =0.72037 (N_1-1)/2$, where the constant was arrived at numerically. One sees that indeed $\f{64}{9 \pi^2} =.7205...$, and hence there is agreement on the principal term, while the $\mathcal{O}(1)$ terms are however different. Indeed Eq.~(\ref{avgnegativity}) agrees very well with numerical results, the differences being indistinguishable from statistical fluctuations.

\subsection{Extreme Value Statistics and entanglement at the critical case}
\label{extstat}

In the critical case when $N_3=4 N_1N_2$, or in terms of the number of qubits $L_1+L_2=L/2-1$, Eq.~(\ref{lognegform}) predicts zero log-negativity as $\tilde{R}=1$. Thus there should
be no NPT states. Numerical calculations however shows that there is a finite fraction of
NPT states. Moreover, and importantly, this is not a finite-size effect. Throughout this section we assume the symmetric case that $N_1=N_2$ so that the semicircle law is valid for the eigenvalue density after PT. The critical case corresponds to $L=4L_1+2$. See Table~\ref{table0030} for a calculation of the percentage of NPT states in several critical cases for 
increasing number of qubits. It is evident that the fraction of NPT states increases with dimensionality. While it is not obvious, it is argued that the fraction of NPT states saturates to a value that is close to $3\%$ and $17\%$ respectively for complex and real states. 

\begin{table}[ht]
\caption{Percentage of NPT states for $L_1=L_2$ and various $L$ for the critical case when $L_1+L_2=L/2-1$.}
 \centering
 \begin{tabular}{|c| c| c| c|}
 \hline \hline 
 $L_1$ & $L$ & \% NPT (Complex states) & \% NPT (Real states)   \\
 \hline 
 1 & 6   & 0.06 $\pm$ 0.008  & 3.18   $\pm$  0.017   \\
 2 & 10  & 1.40  $\pm$ 0.036  & 7.82   $\pm$  0.085   \\
 3 & 14  & 1.92 $\pm$ 0.065  & 11.18  $\pm$ 0.121   \\
 4 & 18  & 2.40  $\pm$ 0.077  & 13.43  $\pm$ 0.161   \\
 5 & 22  & 2.60  $\pm$ 0.145  & 15.17  $\pm$ 0.35    \\
 \hline
 \end{tabular}
 \label{table0030}
 \end{table}
On the other hand when in the neighborhood of criticality (in terms of the number of qubits), $L=4L_1+1$ and $L=4L_1+3$, the percentage of NPT states rapidly 
increases to $100\%$ and decreases to $0 \%$ respectively, 
see Table~\ref{table0029}.  Also one notes that in these cases results have been presented for real states, as the fraction of NPT states are still significant for small $L$ and the numbers are reliable. 

\begin{table}[ht]
\caption{Percentage of NPT states for $L_1=L_2$ and various $L$ (Real states).}
 \centering
 \begin{tabular}{|c| c| c| c| c|}
 \hline \hline 
 $L_1$ & $L=4L_1+1$ & \% NPT & $L=4L_1+3$ & \% NPT  \\
 \hline 
 1 & 5    & 25.39           &  7   & $4.4\times 10^{-2}$   \\
 2 & 9    & 96.82           &  11  & $8.3\times 10^{-5}$  \\
 3 & 13   & $\approx 100 $  &  15  & $ < 10^{-5}$  \\
 4 & 17   &  $\approx 100 $ &  19  & $\approx 0$\\
 5 & 21   &  $\approx 100 $ & 23  & $\approx 0$  \\
 \hline
 \end{tabular}
 \label{table0029}
 \end{table}

As mentioned earlier, for the critical case the radius of the semicircle is such that the hard lower limit is exactly at zero, that is the scaled radius $\tilde{R}=1$. However it is also well known that there is a tail to the semicircle in which the extreme eigenvalues lie. The entire tail is then responsible for the existence of NPT states at criticality. If we are interested in the fraction of NPT states, this is the fraction of states such that $\mu_{min}$, the minimum eigenvalue after PT, is less than $0$. At criticality therefore it is a problem in the theory of extremes. In the absence of a more elaborate random matrix model, we can continue to use the simple model introduced earlier and see how it fares, as the theory of extremes is well-developed for the Gaussian ensembles.  

For $N \times N$ GUE matrices, the diagonal and off-diagonal elements ((both real and imaginary parts) 
are drawn from the normal distributions $\mathcal{N}(0,\sigma^2)$ and $\mathcal{N}(0,\sigma^2/2)$ respectively. The limits of the semicircle are 
$\pm 2 \sigma \sqrt{N}$. While most of the eigenvalues lie in this range, some do not. The problem of estimating the number of eigenvalues outside of this range has been studied for long, 
for example see \cite{Ullah83}. The result about the largest eigenvalue distribution is now stated for $\sigma=1$. If $\lambda_{max}$ is the largest eigenvalue, then 
\beq
x= \dfrac{\lambda_{max} - 2 \sqrt{N}}{N^{-1/6}}
\eeq
has a limiting distribution for large $N$ that is not one of the classical extreme value distributions, but is the Tracy-Widom distribution~\cite{Tracy1,Tracy2}. Thus the $\mbox{Prob}[\lambda_{max} \le x] \rightarrow F_{2}(x)$ and the probability density of the scaled
variable $x$ is $dF_{2}(x)/dx$, where $F_2(x)$ is obtained from a solution of the Painlev\'{e}-II equation. See for example \cite{Edelman05,Edelacta05} for details of a numerical procedure that enables this.
 
\begin{figure}
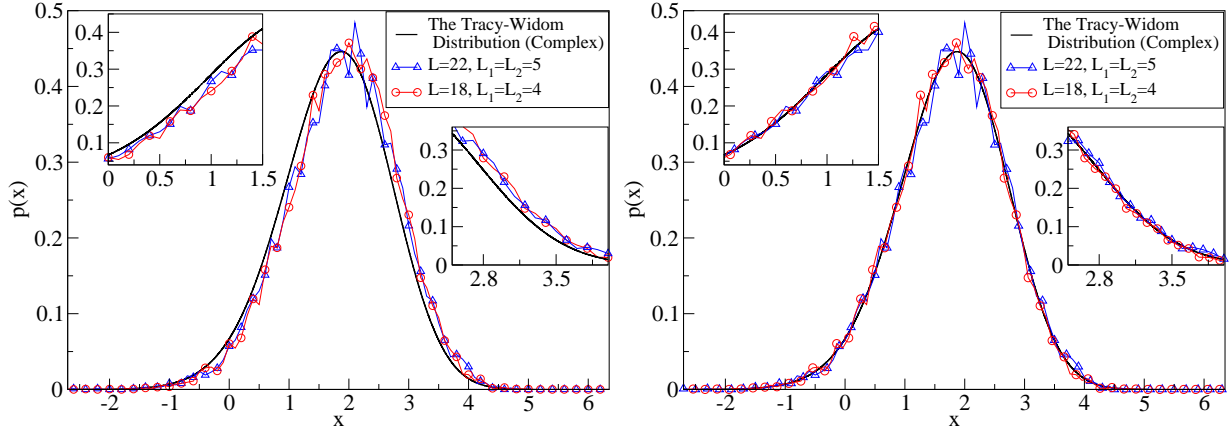

\includegraphics[width=0.49\linewidth,clip]{new10extmin.eps}
\includegraphics[width=0.49\linewidth,clip]{new10extmincorr.eps}   
\caption{(Color online) Distribution of the minimum eigenvalue of $\rho_{12}^{T_2}$ for various cases of $L_1=L_2$ and $L$. The right panel shows the result on using the shift which is determined numerically. This corresponds to complex states after rescaling as given in Eq.~(\ref{scaling}).}
\label{TWidom}
\end{figure}

 Applying the above to the model in Eq.~(\ref{model}), we need to take into account the shifted center and the appropriate variance of the 
 elements of the random matrix $A$. Also we need to consider that we are interested in the minimum rather than the maximum, which is fixed easily as the density of the eigenvalues is symmetric about $0$. Thus the density of the minimum eigenvalue is $p(x)=-dF_2(-x)/dx$. 
 Since $\br \mbox{tr} (A^2) \kt = N^2 \sigma^2$, using Eq.~(\ref{R}) gives the variance of the diagonal elements of $A$ to be $\sigma^2=1/(N^2 N_3)$. Thus we need to consider $(\mu-1/N)\times N \sqrt{N_3}$ as the eigenvalue for a corresponding zero centered GUE with a unit variance for its diagonal elements. Thus the appropriate variable for the minimum eigenvalue after PT is
 \beq
 x=\left( \sqrt{N_3}(N\mu_{min}-1)+ 2 \sqrt{N}\right) N^{1/6}.
 \label{scaling}
 \eeq
As we are especially interested here in the critical case when $N_3=4 N=4 N_1N_2$, the variable $x$ is simply $\sqrt{N_3} N^{7/6} \mu_{min}=2 N^{5/3} \mu_{min}$. 
 Thus the fraction of NPT states, say $f_{NPT}$ is simply the area under the universal Tracy-Widom density, corresponding to $x<0$ keeping in mind that we are now dealing with the minimum eigenvalue. Thus, the simple RMT model for the matrix after PT results in the estimate that 
 \beq
 f_{NPT}=1-F_2(0).
 \eeq
 Note that this is just a number (independent of matrix dimensions that are assumed to be large) that is numerically found to be $\approx .03$. For the case when $L=22, L_1=L_2=5$ qubits we find numerically that there are $2.58\%$ of states that are NPT, thus there is reasonable agreement.  Figure (\ref{TWidom}) shows the distribution of $x$ for two instances of critical dimensions. As the inset indicates, clearly there is a shift from the Tracy-Widom distribution. Indeed the limitations of the model of the PT as a GUE member is reflected in the statistics of the extremes in this way. 
 
\begin{figure}
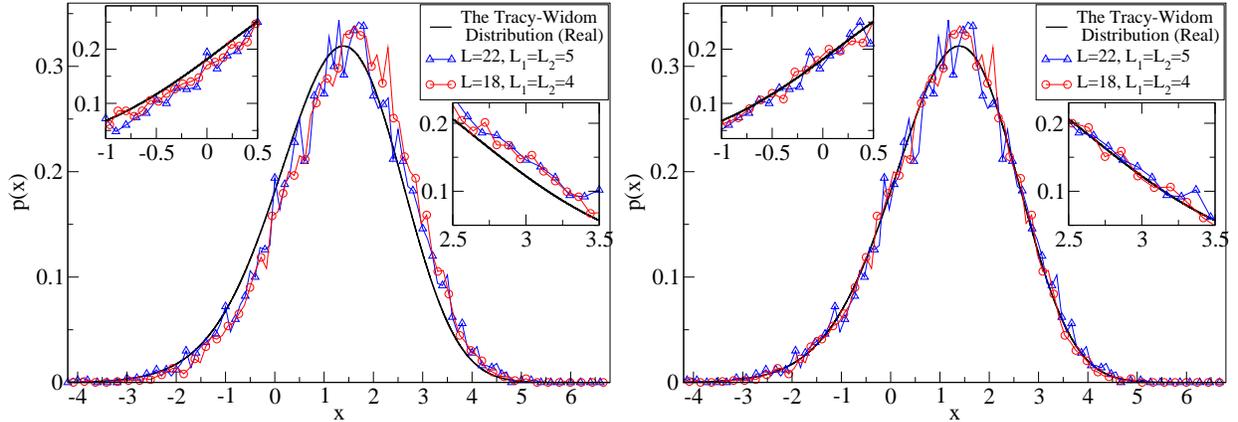

\includegraphics[width=0.49\linewidth,clip]{realextmin.eps}
\includegraphics[width=0.49\linewidth,clip]{realextmincorr.eps}   
\caption{ (Color online) Same as previous figure, but for the case of real states.}
\label{TWidomReal}
\end{figure}

 One needs to add an additional shift for there to be a good match with the Tracy-Widom distribution. A numerically determined shift is applied to
 the two cases and the result is shown in the right panels of Figs.~(\ref{TWidom},\ref{TWidomReal}). The shift is a positive number $s$ such that $x+s$ is given by R.H.S. of Eq.~(\ref{scaling}).
  If the shifted distribution is used for the cases when  ($L=18, L_1=L_2=4$) and ($L=22, L_1=L_2=5$), there are $\approx 2.4 \%$ and $\approx 2.5\%$ of NPT states, in closer agreement with numerical simulations.
 Note that the shift will result in a smaller area as the right end of the integration is moved from zero to $-s$. This shift gets smaller for larger dimensionality and  the fraction of NPT states seems to approach the fraction for the unshifted distribution.
 
Whereas in this section the primary case of complex density matrices and the GUE has been considered, the real case is of considerable interest as well. Prevalence
of time-reversal symmetry in many systems makes the real case important wherein the 
density matrix on PT is modeled by matrices from the GOE.  While there are no essential differences in the density of states after PT, both being close to the Wigner semicircle, they have very distinct distributions for the extreme eigenvalues. As a consequence, at the critical case, the fraction of NPT states is considerably higher for real states. It turns out that the correct scaling for the real case is the same as that in Eq.~(\ref{scaling}) and the distribution of the smallest eigenvalue after PT is shown in Fig.~(\ref{TWidomReal}) for the same dimensions as for the complex case. It is clear from this that the fraction of NPT states is indeed larger, and is $\approx 16.8\%$. In the real case too a shift is needed for good agreement with the relevant Tracy-Widom distribution, 
which is also written in terms of solutions to the Painlev\'{e}-II equation \cite{Tracy1,Tracy2}.
From numerical simulations for the cases when ($L=18, L_1=L_2=4$) and  ($L=22, L_1=L_2=5$), the fraction of NPT states is $13.4\%$ and $15.2\%$ respectively, approaching the $16.8\%$. If the shift is incorporated and then the fraction is calculated there is good agreement even for finite $L$. Thus for example in the case when ($L=14, L_1=L_2=3$) numerical simulations result in $11.16 \%$ of NPT states, while the shift adjusted area under the Tracy-Widom density gives $11.58\%$.

\subsubsection*{Average log-negativity at the critical case}
\label{logcritical}

Using the asymptotic Wigner semicircle results in zero log-negativity, yet there is still
a fraction of NPT states due to the smallest eigenvalues in the tail.  Although the percentage of NPT states can be quite high the log-negativity of the entanglement is small.  Numerical simulations indicate that for at least the dimensions that have been considered here, if there are any negative eigenvalues at all on PT, there is only one. Thus the smallest eigenvalue $\mu_{min}$ almost wholly controls the entanglement in the critical case. Assuming that this is the case gives
\[ 
E_{LN}=\log\big(\sum_i |\mu_i|\big)= \log\big(1-2 \sum_{i; \mu_i<0} \mu_i \big) \approx -2 \mu_{min}
\]
Thus the average log-negativity at critical dimensions is given by 
\beq
\langle E_{LN} \rangle_M \approx -2 \langle \mu_{min} \Theta(-\mu_{min})\kt  = \frac{2}{\sqrt{N_3} N^{7/6}} \int_{-\infty}^{-s} -(x+s) p(x) dx \sim N^{-5/3}, 
\label{avelogtracy}
\eeq
where $\Theta(x)$ is the Heaviside step function and $s>0$ is the numerically determined shift. The final estimate follows from the condition of criticality that $N_3=4 N_1N_2=4N$. The Tables-\ref{criticallogneg2} and \ref{criticallogneg1} show how well this estimates
the average log-negativity in three cases for both complex and real states, respectively. One sees that the real states have a larger entanglement or log-negativity in agreement with their also having a larger fraction of NPT states.

 \begin{table}[ht]
\caption{Average log-negativity for $L_1=L_2$ and various $L$ for the critical case (complex).}
 \centering
 \begin{tabular}{|c| c| c| c| c| c| c| c| c|}
 \hline \hline 
 $L_1$ & $L=4L_1+2$ & Numerical $\langle E_{LN} \rangle$ & $\langle E_{LN} \rangle$ using Eq.(\ref{avelogtracy})   \\
 \hline 
 3 & 14   & $7.28 \times  10^{-6} $ & $8.39 \times  10^{-6} $    \\
 4 & 18   & $9.28 \times  10^{-7} $ & $8.95 \times  10^{-7} $    \\
 5 & 22   & $9.47 \times  10^{-8} $ & $9.79 \times  10^{-8} $    \\
 \hline
 \end{tabular}
 \label{criticallogneg2}
 \end{table}

\begin{table}[ht]
\caption{Average log-negativity for $L_1=L_2$ and various $L$ for the critical case (real).}
 \centering
 \begin{tabular}{|c| c| c| c| c| c| c| c| c|}
 \hline \hline 
 $L_1$ & $L=4L_1+2$ & Numerical $\langle E_{LN} \rangle$ & $\langle E_{LN} \rangle$ using Eq.(\ref{avelogtracy})   \\
 \hline 
 3 & 14   & $7.62 \times  10^{-5} $ & $8.26 \times  10^{-5} $    \\
 4 & 18   & $9.41 \times  10^{-6} $ & $9.51 \times  10^{-6} $    \\
 5 & 22   & $1.13 \times  10^{-6} $ & $1.06 \times  10^{-6} $    \\
 \hline
 \end{tabular}
 \label{criticallogneg1}
 \end{table}

\section{Entanglement amongst three coupled kicked rotors}
\label{stdmap}

To study the applicability of the results above to a dynamical system, this section studies a Hamiltonian system of three coupled and kicked quantum rotors or standard maps. The quantum standard map is one of the most important paradigmsl of 
quantum chaos \cite{Chirikovbook} and has been used extensively from early on \cite{CasatiFord} to study various phenomena such as dynamical localization 
\cite{Izrailev90}. There have been experimental realizations of the quantum standard map using cold atoms where dynamical localization in the momentum has 
been observed. 
Two coupled quantum standard maps were used to study entangling power of quantum chaos \cite{Arulentpow}. More recently, there have been studies of three-dimensional (3D) kicked rotors \cite{Wang3drotor} and many interacting kicked rotors \cite{Lakshminarayan}. 

A single classical standard map on the unit torus is given by the equations
\begin{eqnarray}
q' &=& q+p' \; (\mbox{mod}\ 1)  \nonumber\\
p' &=& p+\frac{K}{2\pi}\sin(2\pi q) \; (\mbox{mod}\ 1)
\end{eqnarray}
which connects phase-space variables $(q_1,p_1)$ just before two consecutive kicks which are separated by a unit of time. The modulo $1$ conditions put the map on a phase space torus, which models conservative systems and Poincar\'{e} surfaces of sections of two-degree-of-freedom systems.  Much is known about the dynamics of the standard map \cite{Liebermanbook}. If $K=0$ then the dynamics is completely integrable. For $K \approx 1$ the last KAM rotational torus breaks, resulting in large scale diffusion in the phase space. For $K<5$ the phase space is a mixed phase space consisting of both regular and chaotic regions.  For $K \gg 5$, the phase space is nearly completely chaotic with only a possibility of finding extremely small islands of regular motion. 

Higher dimensional and coupled standard maps have been previously studied also because new phenomena such as Arnold diffusion arise \cite{Reichlbook}.
The classical coupled maps that are studied in this paper are given by the following canonical or symplectic transformation:
\beq
\begin{split}
q_{i} ' &= q_i+p_{i} '  \; (\mbox{mod} 1) \\
p_{i}' &= p_i+\frac{K_i}{2\pi}\sin(2\pi q_i)+\sum_{j \ne i} \frac{b_{i,j}}{2\pi}\sin[2\pi(q_i+q_j)]  
  \; (\mbox{mod} 1) 
 \label{3classicalmap} 
\end{split}  
\eeq
where the $K_i$'s are parameters for respective maps and $b_{i,j}$ $(i \ne j,\; b_{j,i}=b_{i,j})$ are the couplings.  Here $i,j=1,2,3$ and there are three coupled rotors, with a single body potential and mutual couplings of two-body interactions. Higher dimensional maps such as these are only poorly understood.  This  six dimensional symplectic map is akin to Poincar\'{e} surfaces of section of $4$-degree of freedom systems.  However for the large parameter values that we have studied the maps are fully chaotic, and one may consider its quantization to be one where RMT will be fully applicable.  The question that is being investigated is the entanglement between any two rotors of this tripartite system as measured by the log-negativity.

The quantum standard map is the unitary operator corresponding to the classical
map. It propagates states from one kick to the next. In the position representation it is \cite{Izrailev90,Arul97} 
\beq
U(n',n;K,N) = \frac{1}{\sqrt{iN}} \exp\left[-iN\frac{K}{2\pi} \cos\left(\frac{2\pi}{N}(n+\alpha)\right)\right] \exp\left[ \frac{i \pi}{N} (n'-n)^2\right]. 
\eeq
The phase space being a torus, the quantum mechanics is on a finite dimensional Hilbert space of dimensionality $N$ which is related to a scaled Planck constant as $N=1/h$. Thus the classical limit
is the large $N$ limit. The position kets are labeled by $n=0,...,N-1$ with eigenvalues $(n+\alpha)/N$.
Phase-space reflection symmetry is governed by $\alpha$ and we use $\alpha=.35$ below to avoid 
having symmetric states.
The unitary operator corresponding to the three coupled standard maps in Eq.~(\ref{3classmap}) is
given in the position representation by $\langle n_1'n_2'n_3'|{\cal{U}} |n_1n_2n_3\rangle =$
\beq
\prod_{i=1}^3 U(n_i',n_i;K_i,N_i) \prod_{j>i}
\exp\left\{-i \sqrt{N_iN_j}\frac{b_{i,j}}{2\pi}\cos\left[2\pi\left(\frac{n_i+\alpha}{N_i}+\frac{n_j+\alpha}{N_j}\right)\right]\right\}.
\label{3classmap}
\eeq
Each of the standard maps have their own dimensionality $N_i$. The effective Planck constant is 
$1/(N_1 N_2 N_3)$.

\begin{figure}
\includegraphics[width=0.49\linewidth,clip]{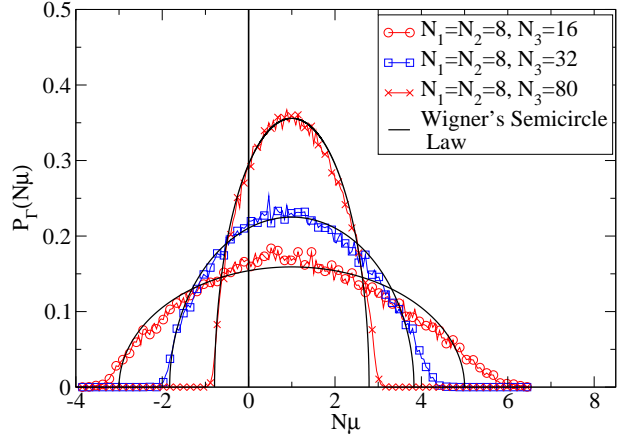}
\caption{(Color online) Density of $\mu$, the eigenvalues after PT, from a thousand eigenstates of ${\cal{U}}$, the quantum map of a set of three coupled standard maps. Dimensions used are $N_1=N_2=8$ and various $N_3$ for parameter set 1, as used in the text.
A vertical line at the origin has been shown to draw attention to the negative part of the spectrum.}
\label{sigma2}
\end{figure}

We study entanglement properties of the eigenstates of ${\cal{U}}$ which are the pure states
of a tripartite system and are the stationary states as far as the quantum map is concerned.
 To be more specific, the entanglement between two rotors, 1 and 2, is studied in these eigenstates using log-negativity. Two parameter sets are used below
 \[
 \begin{split}
 \mbox{Para. Set 1}:&  \;\; (K_1=8, \, K_2=7, \, K_3=6, \, b_{1,2}=1.60, \, b_{1,3}=1.51, \, b_{2,3}=1.42) \\
\mbox{Para. Set 2}:& \;\;  (K_1=15,\,  K_2=14, \, K_3=13, \, b_{1,2}=2.60,\, b_{1,3}=2.51,\,  b_{2,3}=2.42).
\end{split}
\]  
Such large values of $K$'s and $b$'s ensures that the individual standard maps are chaotic and strongly coupled with each other. Also 
the parameters within each set are chosen to be different to break any permutation symmetry effects.
Using 1000 eigenstates of ${\cal{U}}$ the density of states of $\rho_{12}^{T_2}$ is shown in Fig.~(\ref{sigma2})) for parameter set $1$, with the $64000$ eigenvalues at one's disposal.  Here we see that the distribution fits reasonably well with that of a corresponding Wigner's semicircle law (see Eq.~(\ref{wignerdist}),  and recall that $R=2/\sqrt{N_1N_2N_3}$, $N=N_1N_2$). There are deviations in the tail regions especially at the large eigenvalues, and agreement between the two distributions improves as $N_3$ increases.
A similar kind of behavior in the density of states of $\rho_{12}^{T_2}$ for the parameter set $2$ was observed and is not presented here. It is perhaps amusing that the Wigner semicircle appears, perhaps for the first time, in the study of a dynamical system, but in the properties of the eigenstates rather than the eigenvalues.

\begin{figure}
\includegraphics[width=0.49\linewidth,clip]{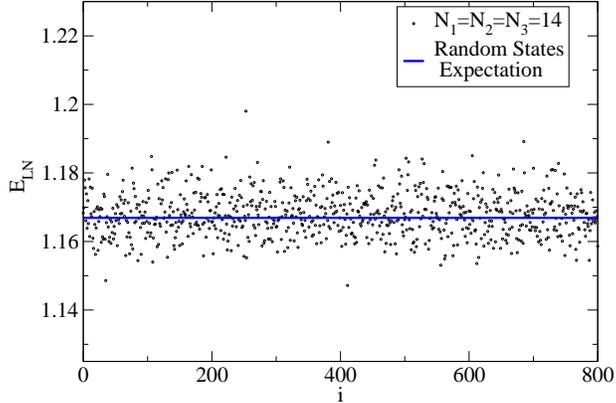} 
\caption{(Color online) Log-negativity ($E_{LN}$) for 800 eigenstates of the three coupled standard maps, the unitary operator $\cal{U}$, for parameter set $1$ (see text). 
The solid horizontal line is the average log-negativity of random states of corresponding dimensions, and is given for all practical purposes by Eq.~(\ref{lognegform}).}
\label{sigma99}
\end{figure}

If $4N_1N_2>N_3$, and a typical random state is NPT, the eigenstates of the coupled standard maps have a log-negativity that is close to that of random states, but consistently slightly larger. See Fig.~(\ref{sigma99}) for the log-negativity between maps 1 and 2 for a sample set of eigenstates. For parameter set 2 the average log-negativity is closer to that of random states compared to parameter set 1 (refer Table \ref{table1}) perhaps reflecting the increased chaos in the classical system, although most of the standard diagnostics of quantum chaos, such as the nearest neighbor spacing statistics do not differentiate between the two sets. 

The increased entanglement, as measured by the log-negativity, for the standard map in comparison to random states,  is consistent with lowered multipartite entanglement between 1,2, and 3, as well as with lowered entanglement between 1+2 and 3. In terms of the monogamy of entanglement, 1 and 2 can be more entangled as they are less entangled with 3, as compared to a typical random state. If we view the third rotor as the environment, it implies a smaller decoherence for the subsystem 1+2.
Thus we can say that the log-negativity in these cases furthers the BGS conjecture that quantum chaotic systems have RMT properties, but at the same time provides rather stringent and new tests for this. This is even more acute in the case of critical dimensions.

\begin{table}[ht]
\caption{Average log-negativity.}
 \centering
 \begin{tabular}{|c| c| c| c| c| c|}
 \hline \hline 
 $N_1$ & $N_2$ & $N_3$ & Para. Set 1 & Para. Set 2 & Real random states\\
 \hline 
 8 & 8 & 32 & 0.3567 &  0.3558 & 0.3491\\
 8 & 8 & 80 & 0.1055 & 0.1054 & 0.1005\\
 10 & 10 &10 & 1.0041 & 1.0035 & 1.0032\\
 12 & 12 &12 & 1.0926 & 1.0922 & 1.0918\\
 14 & 14 &14 & 1.1678 & 1.1676 & 1.1669\\
 \hline
 \end{tabular}
 \label{table1}
 \end{table}

\begin{table}[ht]
\caption{Average log-negativity for critical cases.}
 \centering
 \begin{tabular}{|c| c| c| c| c| c|}
 \hline \hline 
 $N_1$ & $N_2$ & $N_3$ & Para. Set 1 & Para. Set 2 & Real random states\\
 \hline 
 4 & 4 & 64 & $4.40 \times 10^{-3}$ & $3.90 \times 10^{-3}$ & $5.28 \times 10^{-4}$\\
 6 & 6 & 144 & $4.85 \times 10^{-4}$ & $4.35 \times 10^{-4}$ & $2.24 \times 10^{-4}$\\
 \hline
 \end{tabular}
 \label{table2}
 \end{table}

\begin{figure}
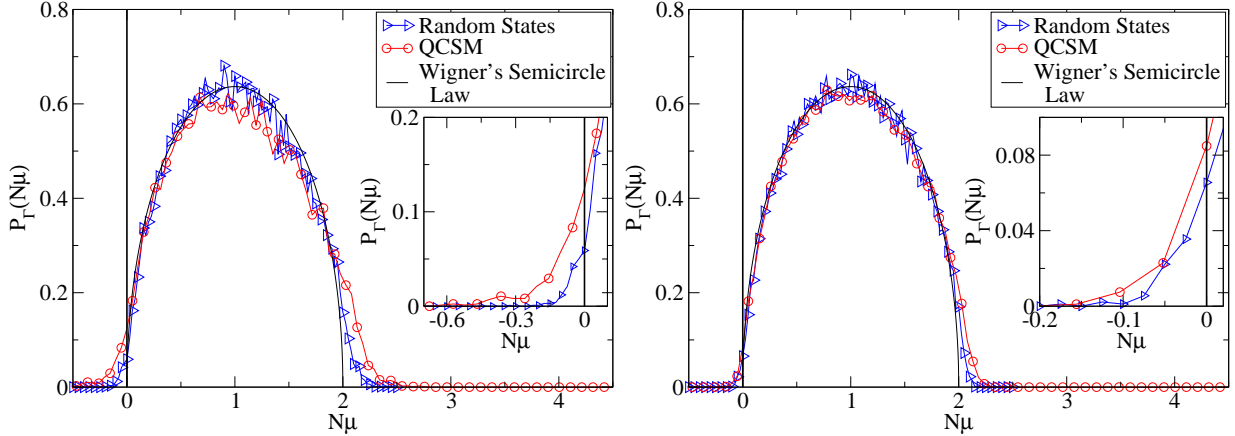

\includegraphics[width=0.49\linewidth,clip]{4464csmrand.eps}
\includegraphics[width=0.49\linewidth,clip]{66144csmrand.eps} 
\caption{(Color online) Density of $\mu$, the eigenvalues after PT, for eigenstates of the three coupled standard maps, for various dimensions and using the parameter set 1 (see text).
Figure on the left corresponds to $N_1=N_2=4$, $N_3=64$ and that on the right $N_1=N_2=6$, $N_3=144$, which are critical cases. The insets show an enlarged view of the region near the origin of the respective figures.
A vertical line at the origin has been shown, as before, to draw attention to the negative part of the spectrum.}
\label{sigma7}
\end{figure}

In the critical cases when $4N_1N_2=N_3$ and the majority of random states are PPT, 
but there is a fraction of NPT states, the average log-negativity for the coupled standard maps is systematically again more than that of the random states of corresponding 
dimensions, see Table \ref{table2}. The distribution of the eigenvalues after PT, near the left tail in Fig.~(\ref{sigma7}) differ from that of random states and are highlighted in the inset of 
this figure. Since the area under the curve for $\mu<0$ is larger for the eigenstates of the coupled standard map than that of random states there is a larger average log-negativity for the former case. Also using parameter set 2 we see that the average log-negativity of the eigenstates of the coupled standard maps tends to that of random states as shown in Table \ref{table2}.

The percentage of NPT states for two critical cases is presented in Table \ref{table29}, where it is 
compared with that of real random states of corresponding dimensions. While the log-negativity is itself larger for the standard maps, the differences are not great. However in terms of the percentage of  NPT states the differences between the dynamical system and the random states are stark.  This data however does show that the percentage of NPT states of coupled standard map eigenstates is closer to that of real random states for parameter set 2 than parameter set 1, again maybe a reflection of increased classical chaos, and that the RMT values may be reached asymptotically. However for finite quantum systems, where other diagnostics indicate agreement with RMT, such tests seem to show still influences of a dynamical  origin.
\begin{table}[ht]
\caption{Percentage of NPT states for critical cases.}
 \centering
 \begin{tabular}{|c| c| c| c| c| c| }
 \hline \hline 
 $N_1$ & $N_2$ & $N_3$ & Para. Set 1 & Para. Set 2&Real random states \\
 \hline 
 4 & 4 & 64 & 29.30\% & 20.37\% & 7.82\%   \\
 6 & 6 & 144 & 23.17\% & 18.85\% & 9.99\%  \\
 \hline
 \end{tabular}
 \label{table29}
 \end{table}

Finally the skewness of the density of states of $\rho_{12}^{T_2}$ of the eigenstates of the quantum standard map $\cal{U}$ is compared 
with that of the analytical formula (Eq.~(\ref{skewform})) for the real case in Table ~(\ref{table22}). One observes that as the 
dimension of the individual standard maps increases this skewness tends to that of corresponding random real states.
When the dimension of the two standard maps $1+2$ is small, and that of third is large the
skewness approaches the random case, indicating once more increased decoherence from the 
third rotor. A more systematic study of the coupled standard maps, for various dynamical regimes
and for other dimensions is postponed. The primary purpose of the present selection is to 
indicate relevant dynamical systems where we may see easily the results on entanglement 
of partial subsystems.

\begin{table}[ht]
\caption{Average skewness.}
 \centering
 \begin{tabular}{|c| c| c| c| c| c| c|}
 \hline \hline 
 $N_1$ & $N_2$ & $N_3$ & CSM Para 1 (using Eq.~(\ref{skewformana})) & Analytical using Eq.~(\ref{skewform})\\
 \hline 
 4 & 4 & 64 & 0.1126                  & $6.25 \times 10^{-2}$  \\
 4 & 4 & 150 & $8.05 \times 10^{-2}$  & $4.08 \times 10^{-2}$  \\
 4 & 4 & 200 & $7 \times 10^{-2}$     & $3.53 \times 10^{-2}$  \\
 8 & 8 & 16 & 0.1150                  & $6.25 \times 10^{-2}$  \\
 8 & 8 & 32 & $8.14 \times 10^{-2}$   & $4.42 \times 10^{-2}$  \\
 8 & 8 & 80 & $5.18 \times 10^{-2}$   & $2.79 \times 10^{-2}$  \\
 6 & 6 & 144 & $5.74 \times 10^{-2}$  & $2.77 \times 10^{-2}$  \\  
 12 & 12 & 12 & $7.81 \times 10^{-2}$ & $4.81 \times 10^{-2}$ \\
 14 & 14 & 14 & $6 \times 10^{-2}$    & $3.81 \times 10^{-2}$ \\
 16 & 16 & 16 & $4.74 \times 10^{-2}$ & $3.12 \times 10^{-2}$ \\
 18 & 18 & 18 & $3.86 \times 10^{-2}$ & $ 2.61 \times 10^{-2}$ \\
 \hline
 \end{tabular}
 \label{table22}
 \end{table}

\section{Summary and conclusions}

This paper has dealt with entanglement amongst two subsystems (say 1 and 2) of random tripartite pure states, using log-negativity as the measure. It is found that the state of a subsystem is typically NPT, and hence entangled, if the number of qubits in it ($L_1+L_2$) is larger than half the total number ($L$). To be precise, the number of qubits in the subsystem should be larger than $L/2-1$, otherwise the state is typically PPT, the critical case being when $L_1+L_2=L/2-1$. It is known that the eigenvalue distribution of the reduced density matrix of a subsystem is given by the Marcenko-Pastur function; but it is found numerically that the same for the reduced density matrix of subsystems after PT is close to the Wigner semicircle law, especially when the number of qubits in the subsystems is not very close to the total number of qubits.  A simple random matrix model, proposed herein, captures both the NPT-PPT transition and the spectral features after PT reasonably well.

An analytical formula for the average log-negativity is derived using the Wigner semi-circle law, which is in good agreement with numerical simulations.
This formula deviates considerably when the number of qubits in the subsystem is  equal to the total number of qubits because the eigenvalue distribution of the reduced density matrix after PT differs from the Wigner semicircle law. In this case, using tools of random matrix theory, an analytical expression for the average log-negativity is given that holds even if the subsystems differ in size. This also generalizes and augments expressions for the average negativity in \cite{Animeshrandom}.
 
An exact expression for the average of the third moment of the reduced density matrix after PT has also been derived, both for the complex and real cases.
This quantity, which is the first moment to deviate from that of the density matrix, is remarkable in possessing permutation symmetry amongst the three subsystems. In fact it is proved that this symmetry is possessed not just on average but by individual states also. Therefore it can be considered as a possible measure of entanglement, especially as it is also a local unitary invariant and is a qudit generalization of the Kempe invariant $I_5$ \cite{Kempe99,Sudbery01}.

Using the Wigner semicircle density, the fraction of NPT states  and hence the average log-negativity at criticality is, zero. However, a small but definite fraction 
of states is found to be NPT. This fraction and the associated entanglement is found by 
using the Tracy-Widom distribution for the extreme eigenvalues of random matrices,
and is in good agreement with numerics, especially since it is observed that in all the cases that we have come across, whenever a state is NPT only one of its eigenvalues is negative. This constitutes perhaps the first use of the well-known Tracy-Widom distribution in quantum information theory.

Finally eigenvectors of three coupled standard maps (or kicked rotors) were studied and compared with that of random real states where the parameters of the map were adjusted such that the classical dynamics is fully chaotic and the quantization preserves time-reversal symmetry. While agreeing for the most part with the results of random states,  deviations are seen prominently at critical dimensions. The deviations are consistent with the dynamical states possessing marginally lower tripartite entanglement than random states. It is interesting that deviations are highlighted in quantities studied here, and hence they provide rather stringent tests of the BGS conjecture that random matrices are models of quantum chaotic systems.

This work suggests several future directions. For example, the joint probability density function of eigenvalues after PT is not known, but lies presumably ``between" the Laguerre and the Gaussian ensembles. Large deviation theory can be used to give estimates of the extremely small fraction of NPT states for $L_1+L_2 < L/2-1$ when the states are dominantly PPT. The comparison with dynamical models
such as spin chains and oscillators will be interesting, especially in regimes where
random matrix theory may not hold. Finally we emphasize the occurrence of the Wigner semi-circle law in quantum chaotic systems, originating not in the eigenvalues, but in the properties of eigenfunctions. 

\appendix
\section{An exact evaluation of $\left \langle \mbox{tr}\left [ \left(\rho_{12}^{T_2}\right)^3 \right] \right \rangle $}

In this appendix an exact evaluation of the ensemble average of the third moment of the PT, the first to depart from that of the density matrix, is calculated. Hence the skewness of the eigenvalue density of $\rho_{12}^{T_2}$ is found.
Here the notations $N_i=2^{L_i}$, $N=N_1N_2$ and $M=N_1N_2N_3$ are used.
For a bipartite partition of a pure state of $L_1+L_2$ qubits and the remaining $L_3=L-L_1-L_2$ qubits, a general state $|\psi \rangle$ is given by 
\begin{eqnarray}
 |\psi \rangle  &=&  \sum_{i=0}^{N-1}\;\; \sum_{n=0}^{N_3-1} a_{in}|i\rangle |n\rangle ,\; 0\leq L_1+L_2 \leq L.
\end{eqnarray}
Hence the reduced density matrix of $L_1+L_2$ qubits ($\rho_{12}$) and $\mbox{tr}\left[\left(\rho_{12}\right)^3\right]$ are given by
\begin{eqnarray}
\begin{split}
(\rho_{12})_{ij} &= \sum_{m=0}^{N_3-1}a_{im} a_{jm}^*,\\
\mbox{tr}\left(\left(\rho_{12}\right)^3\right) &= \sum_{i,j,k=0}^{N-1} \;\;\sum_{m,n,p=0}^{N_3-1}\; a_{im}a_{jm}^*a_{jn}a_{kn}^*a_{kp}a_{ip}^*.
\end{split}
\label{trrho3}
\end{eqnarray}

After PT $   \left(\rho_{12}^{T_2}\right)_{\tilde{i}\tilde{j}} = (\rho_{12})_{ij}$ where 
\begin{eqnarray}
\label{definegij}
\begin{split}
\tilde{i} :=g(i,j)&= i-\mbox{mod}(i,N_2)+\mbox{mod}(j,N_2),\\
\tilde{j} :=g(j,i)&= j-\mbox{mod}(j,N_2)+\mbox{mod}(i,N_2).
\end{split}
\end{eqnarray}
The function $(i,j) \mapsto (\tilde{i},\tilde{j})=(g(i,j),g(j,i))$ is bijective and is its own inverse, since
performing PT twice keeps elements of $\rho_{12}$ unchanged. This implies that $i=g(\tilde{i},\tilde{j})$ and $j=g(\tilde{j},\tilde{i})$, {\it i.e.}  $\left(\rho_{12}^{T_2}\right)_{ij} =(\rho_{12})_{\tilde{i}\tilde{j}}.$ Thus we state the following simple conclusions are useful lemmas:

 Lemma 1 $\tilde{i}=\tilde{j}$ iff $i=j$. 
 
 Lemma 2 $g(i,i)=i$.

Lemma 3  If $(i,j) \mapsto (g(i,j),g(j,i))$ and $(i',j') \mapsto (g(i',j'),g(j',i'))$, and $j\ne j'$ $(i\ne i')$, then $g(i,j) \ne g(i',j')$ $(g(j,i)\ne g(j',i'))$. In words, elements that differ in column (row) position get mapped after PT to positions that differ in row (column). 

The expression for $\mbox{tr}[(\rho_{12}^{T_2})^3]$ using Eqs.~(\ref{trrho3}) and the above function is 
\begin{eqnarray}
\mbox{tr}[(\rho_{12}^{T_2})^3] = \sum_{i,j,k=0}^{N-1} \;\;\sum_{m,n,p=0}^{N_3-1}\; a_{\tilde{i_1}m}a_{\tilde{j_1}m}^*
a_{\tilde{j_2}n}a_{\tilde{k_2}n}^*a_{\tilde{k_3}p}a_{\tilde{i_3}p}^*
\label{rhoptexp}
\end{eqnarray}
where $\tilde{i_1}=g(i,j)$, $\tilde{j_1}=g(j,i)$, $\tilde{j_2}=g(j,k)$, $\tilde{k_2}=g(k,j)$, $\tilde{k_3}=g(k,i)$ and 
$\tilde{i_3}=g(i,k)$.
Using this one obtains the ensemble average $\langle\mbox{tr}[(\rho_{12}^{T_2})^3]\rangle$ as follows. Here the fact that after averaging only even powered terms will be nonzero and odd powered terms will be zero is used.
It can be seen in Eq.~(\ref{rhoptexp}) that there are three possible cases for $m$, $n$ and $p$, namely $m\neq n\neq p$, $m=n\neq p$ (which is the same as $m\neq n=p$ and $m\neq p=n $) and $m=n=p$. In each of these cases the number of terms that do not vanish on averaging is first calculated, the last case requiring a somewhat detailed analysis.

\bc
{\bf Case: }$\mathbf {m\neq n\neq p}$
\ec
In this case, the only possible non-vanishing term after averaging is a product of three unequal quadratic terms {\it i.e.} $\tilde{i_1}=\tilde{j_1}$,
$\tilde{j_2}=\tilde{k_2}$ and $\tilde{k_3}=\tilde{i_3}$, which implies that  $g(i,j)=g(j,i)$, $g(j,k)=g(k,j)$ and $g(k,i)=g(i,k)$. As a consequence of the Lemma 1 above, one then obtains that $i=j=k$. Thus in this case there are exactly $N$ non-vanishing terms on averaging 
Eq.~(\ref{rhoptexp}), each of the form $|a_{im}|^2 |a_{in}|^2 |a_{ip}|^2$.

\bc
{\bf Case: }$\mathbf {m=n\neq p}$
\ec

Now there are two possible types of non-vanishing terms, one a product of three unequal quadratic terms and the other a product of one quadratic and one quartic term. In the former possibility, $\tilde{k_3}=\tilde{i_3}$,
$\tilde{i_1}=\tilde{k_2}$, $\tilde{j_1}=\tilde{j_2}$ and $\tilde{i_1} \neq \tilde{j_1}$ {\it i.e.} $g(k,i)=g(i,k)$, $g(i,j)=g(k,j)$, 
$g(j,i)=g(j,k)$ and $g(i,j)\neq g(j,i)$. Again using the Lemma 1, this implies that $i=k\neq j$. Thus there are $N(N-1)$ non-vanishing terms on averaging, each of the form $|a_{im}|^2 |a_{jm}|^2 |a_{ip}|^2$.

In the later case, $\tilde{k_3}=\tilde{i_3}$ and $\tilde{i_1}=\tilde{j_1}=\tilde{j_2}=\tilde{k_2}$ 
{\it i.e.} $g(k,i)=g(i,k)$ and $g(i,j)=g(j,i)=g(j,k)=g(k,j)$ which implies $i=k=j$. Thus there are $N$ non-vanishing terms of the form $|a_{im}|^4  |a_{ip}|^2$.

\bc
{\bf Case: }$\mathbf {m=n= p}$
\ec
In this case there are three possible types of non-vanishing terms:
(1) one sextic term, (2) one quadratic and one quartic term and (3) three unequal quadratic terms.
\begin{enumerate}
\item One sextic term. This occurs when $\tilde{i_1}=\tilde{j_1}=\tilde{j_2}=\tilde{k_2}=\tilde{k_3}=\tilde{i_3}$, that is $g(i,j)=g(j,i)=g(j,k)=g(k,j)=g(k,i)=g(i,k)$. This in turn implies that $i=j=k$. Thus there are $N$ non-vanishing terms of the form $|a_{im}|^6$.

\item One quadratic and one quartic term. Corresponding to such terms there are two cases.
\begin{enumerate}

\item $\tilde{i_1}=\tilde{k_2}$, $\tilde{j_1}=\tilde{j_2}=\tilde{k_3}=\tilde{i_3}$ and $\tilde{i_1} \neq \tilde{j_1}$. 

In terms of function $g$ this condition is $g(i,j)=g(k,j)$, $g(j,i)=g(j,k)=g(k,i)=g(i,k)$ and $g(i,j)\neq g(j,i)$.
This implies, using the Lemmas above, that  $i=k$ and $g(j,k)=k$ but $k \neq j$. Thus $j-\mbox{mod}(j,N_2)+\mbox{mod}(k,N_2)=k$, i.e. 
$j-\mbox{mod}(j,N_2)=k-\mbox{mod}(k,N_2)$ and $k \neq j$. In other words, $j$ and $k$ have to be distinct, but have the same quotient on division by $N_2$.
Here $j$ and $k$ takes values from $0$ to $N-1$ where $N=N_1N_2$. To find the number of $(j,k)$ pairs that satisfy these conditions, imagine dividing $N$ into $N_1$ intervals each of length $N_2$. If $j$ is selected from the $N$ possible numbers, this also fixes one such interval. The number $k$ must necessarily be in this interval, but must not be $j$, which gives a choice of multiplicity $(N_2-1)$.Thus there are $N(N_2-1)$ non-vanishing terms 
of the form $|a_{jm}|^4 |a_{km}|^2$.

\item $\tilde{i_1}=\tilde{i_3}$, $\tilde{j_1}=\tilde{j_2}=\tilde{k_2}=\tilde{k_3}$ and $\tilde{i_1} \neq \tilde{j_1}$.

In terms of the function $g$ this condition is $g(i,j)=g(i,k)$, $g(j,i)=g(j,k)=g(k,j)=g(k,i)$ and $g(i,j)\neq g(j,i)$. Using the Lemmas above implies that $j=k$ and $g(j,i)=j$ but $i \neq j$.
Thus $j-\mbox{mod}(j,N_2)+\mbox{mod}(i,N_2)=j$ i.e. $\mbox{mod}(j,N_2)=\mbox{mod}(i,N_2)$ and $i \neq j$. In words, one must count the number of distinct pairs of integers having the same remainder on division by $N_2$. This is easily seen from an argument similar to that in the previous paragraph to be $N(N_1-1)$, each corresponding to a non-vanishing term of the form $|a_{im}|^2|a_{jm}|^4$. 
\end{enumerate}

Combining these two cases and including the cyclic permutation of $(i,j,k)$, the total number of non-vanishing terms of this kind are $3N(N_1+N_2-2)$. 

We note in parenthesis that this is the first instance that the counting is different from that 
for evaluating $\langle\mbox{tr}[(\rho_{12})^3]\rangle$ in which one has the condition 
$i=j\neq k$ and its cyclic permutations, implying $3N(N-1)$ non-vanishing terms.

\item Three unequal quadratic terms.
For three unequal quadratic terms there are fifteen possible cases, of which only seven are distinct up to cyclic permutation of indices $(i,j,k)$. These are listed in Table (\ref{conditions}), and subsequently analyzed. 
\begin{table}[ht]
\caption{}
 \centering
 \begin{tabular}{|c| c|}
 \hline
(a) & $(\tilde{i_1}=\tilde{j_1}) \neq (\tilde{j_2}=\tilde{k_2}) \neq (\tilde{k_3}=\tilde{i_3})$ \\
(b) & $(\tilde{i_1}=\tilde{j_1}) \neq (\tilde{j_2}=\tilde{k_3}) \neq (\tilde{k_2}=\tilde{i_3})$ \\
(c)& $(\tilde{i_1}=\tilde{j_1}) \neq (\tilde{j_2}=\tilde{i_3}) \neq (\tilde{k_2}=\tilde{k_3})$ \\
(d) & $(\tilde{i_1}=\tilde{j_2}) \neq (\tilde{j_1}=\tilde{k_3}) \neq (\tilde{k_2}=\tilde{i_3})$ \\
(e) &  $(\tilde{i_1}=\tilde{j_2}) \neq (\tilde{j_1}=\tilde{i_3}) \neq (\tilde{k_2}=\tilde{k_3})$\\
(f) & $(\tilde{i_1}=\tilde{k_2}) \neq (\tilde{j_1}=\tilde{k_3}) \neq (\tilde{j_2}=\tilde{i_3})$ \\
(g) & $(\tilde{i_1}=\tilde{i_3}) \neq (\tilde{j_1}=\tilde{j_2}) \neq (\tilde{k_2}=\tilde{k_3})$  \\ 
 \hline
\end{tabular}
\label{conditions}
\end{table}
\begin{enumerate}
\item $(\tilde{i_1}=\tilde{j_1}) \neq (\tilde{j_2}=\tilde{k_2}) \neq (\tilde{k_3}=\tilde{i_3})$.

In terms of function $g$ this condition is
$\left(g(i,j)=g(j,i)\right) \neq \left(g(j,k)=g(k,j)\right) \neq \left(g(k,i)=g(i,k)\right)$. 
This gives $i=j$, $j=k$, $i=k$, $i \neq j$, $j \neq k$ and $i \neq k$, which are incompatible conditions and hence the number of terms of this kind is zero. The multiplicity of this case under
cyclic permutation of the labels $(i,j,k)$ is $1$.

\item$(\tilde{i_1}=\tilde{j_1}) \neq (\tilde{j_2}=\tilde{k_3}) \neq (\tilde{k_2}=\tilde{i_3})$.

In terms of function $g$ this condition is
$\left(g(i,j)=g(j,i)\right) \neq \left(g(j,k)= g(k,i)\right) \neq \left(g(k,j)=g(i,k)\right)$.
The first equality gives  $i=j$, and the inequation $g(j,k)\ne g(k,j)$ implies that $j \ne k$.
However the second equality (using $i=j$) implies that $i=k$. Thus these conditions are 
incompatible and the number of terms of this kind is zero.  The multiplicity of this case under
cyclic permutation of the labels $(i,j,k)$ is $3$.

\item $(\tilde{i_1}=\tilde{j_1}) \neq (\tilde{j_2}=\tilde{i_3}) \neq (\tilde{k_2}=\tilde{k_3})$.

 In terms of function $g$ this condition is
$\left(g(i,j)=g(j,i)\right) \neq \left(g(j,k)= g(i,k)\right) \neq \left(g(k,j)=g(k,i)\right)$.
This first equality gives $i=j$,  while the inequation $g(i,k) \ne g(k,i)$ implies that $i\neq k$.
Further $g(i,j)=g(i,i)=i \ne g(i,k) \Rightarrow \mbox{mod}(i,N_2) \neq \mbox{mod}(k,N_2)$ and $i \ne g(k,i) \Rightarrow i-\mbox{mod}(i,N_2) \neq k-\mbox{mod}(k,N_2)$ i.e.
remainder and quotient of $i$ and $k$, under division by $N_2$, are not same. Again dividing $N$ into $N_1$ intervals each of length $N_2$ gives 
$N(N-N_1-N_2+1)$ as the number of possible triples $(i,j,k)$ that satisfy these conditions. 
 The multiplicity of this case under
cyclic permutation of the labels $(i,j,k)$ is $3$.
 
\item $(\tilde{i_1}=\tilde{j_2}) \neq (\tilde{j_1}=\tilde{k_3}) \neq (\tilde{k_2}=\tilde{i_3})$.

In terms of function $g$ this condition is
$\left(g(i,j)=g(j,k)\right) \neq \left(g(j,i)=g(k,i)\right) \neq \left(g(k,j)=g(i,k)\right)$.
The inequations and the Lemma 1 above imply that $i \ne j \ne k$. It is then not hard to 
see that the equalities are incompatible with this condition. For instance $(i,j) \mapsto (g(i,j),g(j,i))$
whereas $(j,k) \mapsto (g(j,k),g(k,j))$, and using Lemma 3 with $(i',j') \equiv (j,k)$, $g(i,j) \ne g(j,k)$, which violates one of the requirements. Thus the number of terms of this kind is zero. 
 The multiplicity of this case under cyclic permutation of the labels $(i,j,k)$ is $3$.

\item $(\tilde{i_1}=\tilde{j_2}) \neq (\tilde{j_1}=\tilde{i_3}) \neq (\tilde{k_2}=\tilde{k_3})$.

 In terms of function $g$ this condition is
$\left(g(i,j)=g(j,k)\right) \neq \left(g(j,i)=g(i,k)\right) \neq \left(g(k,j)=g(k,i)\right)$. An analysis very similar to the previous case shows that these conditions are incompatible too.  The multiplicity of this case under
cyclic permutation of the labels $(i,j,k)$ is $3$.

\item $(\tilde{i_1}=\tilde{k_2}) \neq (\tilde{j_1}=\tilde{k_3}) \neq (\tilde{j_2}=\tilde{i_3})$.

In terms of function $g$ this condition is
$\left(g(i,j)=g(k,j)\right) \neq \left(g(j,i)=g(k,i)\right) \neq \left(g(j,k)=g(i,k)\right)$. The inequations again imply that $i \ne j\ne k$, however the Lemma's do not lead to incompatible conditions.
\[ 
\begin{split}
g(i,j)&=g(k,j) \Rightarrow i-\mbox{mod}(i,N_2)=k-\mbox{mod}(k,N_2)\\ g(j,i)&=g(k,i) \Rightarrow j-\mbox{mod}(j,N_2)=k-\mbox{mod}(k,N_2)\\
g(j,k)&=g(i,k)\Rightarrow i-\mbox{mod}(i,N_2)=j-\mbox{mod}(j,N_2).\end{split} \]
Thus $i$, $j$, $k$ have the same quotient on division by $N_2$, and as they are all distinct they
have different remainders.
To find the number of triples $(i,j,k)$ that satisfy these conditions, once more divide an interval of length $N$ into $N_1$ intervals of length $N_2$. One can select $i$ in $N$ possible ways, 
which fixes the quotient on division by $N_2$. The integers $j$ and $k$ must then be one of the possible $N_2-1$ numbers, without also being equal. Hence the number of terms of this kind is $N(N_2-1)(N_2-2)$.  The multiplicity of this case under
cyclic permutation of the labels $(i,j,k)$ is $1$.

\item  $(\tilde{i_1}=\tilde{i_3}) \neq (\tilde{j_1}=\tilde{j_2}) \neq (\tilde{k_2}=\tilde{k_3})$.

 In terms of function $g$ this condition is
$\left(g(i,j)=g(i,k)\right) \neq \left(g(j,i)=g(j,k)\right) \neq \left(g(k,j)=g(k,i)\right)$.
 A similar analysis as for the previous case shows that $i$, $j$ and $k$ are distinct but they have a common remainder on division by $N_2$. Thus there will be $N(N_1-1)(N_1-2)$ number of non-vanishing terms in this case.  The multiplicity of this case under
cyclic permutation of the labels $(i,j,k)$ is $1$.

 \end{enumerate}

Thus the number of terms with a product of three distinct quadratics, including the multiplicities is $3N(N-N_1-N_2+1)+N(N_1-1)(N_1-2)+N(N_2-1)(N_2-2) $.
In contrast the number of such terms in the evaluation of  $\langle\mbox{tr}[(\rho_{12})^3]\rangle$ is $N(N-1)(N-2)$. In general note that one can recover results for the density matrix prior to PT from those after PT by replacing $N_2 \rarrow 1$ and $N_1 \rarrow N$. Thus the results for PT present a particular generalization.

\end{enumerate}

The exact RMT ensemble average values \cite{Ullah} for the case of  complex states are stated below for convenience:
\[ \begin{split} 
\langle|a_{in}|^6 \rangle = \df{6}{M(M+1)(M+2)},
\langle|a_{in}|^4|a_{jn}|^2\rangle =\langle|a_{in}|^4|a_{im}|^2\rangle =\df{2}{M(M+1)(M+2)},&\\
\langle|a_{in}|^2 |a_{jm}|^2 |a_{kp}|^2\rangle=\langle|a_{in}|^2 |a_{im}|^2 |a_{kp}|^2\rangle=
\langle|a_{in}|^2 |a_{im}|^2 |a_{ip}|^2\rangle=
\langle|a_{in}|^2 |a_{jn}|^2 |a_{kp}|^2\rangle=&\\ \langle|a_{in}|^2 |a_{jn}|^2 |a_{ip}|^2\rangle=
\langle|a_{in}|^2 |a_{jn}|^2 |a_{jp}|^2\rangle=\langle|a_{in}|^2 |a_{jn}|^2 |a_{kn}|^2\rangle=\df{1}{M(M+1)(M+2)}&.\\
\end{split} \]
These averages are multiplied by the number of non-vanishing terms and by the respective multiplicity for $m$, $n$ and $p$ (for $m\ne n \ne p$ it is $N_3(N_3-1)(N_3-2)$, for $m=n \ne p$ it is $3N_3(N_3-1)$ and for $m=n=p$ it is simply $N_3$), and added together. Use is made 
of $N N_3=N_1N_2N_3=M$ and a straightforward simplification of 22 terms results in 
significant cancellations, leaving just 4 terms finally. This results in 
\begin{equation}
\langle\mbox{tr}\big(\rho_{12}^{T_2}\big)^3\rangle =\dfrac{N_1^2+N_2^2+N_3^2+3N_1N_2N_3}{
(N_1N_2N_3+1)(N_1N_2N_3+2)},
\label{tr3cmpx}
\eeq
with the remarkable permutation symmetry evidently displayed.
 
A  similar analysis can be done for the case of averaging only over real states. The counting 
remains identical to the complex case, while the averages differ as:
\[ \begin{split} 
\langle a_{in} ^6 \rangle = \df{15}{M(M+2)(M+4)},
\langle a_{in}^4 a_{jn}^2\rangle =\langle a_{in}^4 a_{im}^2\rangle =\df{3}{M(M+2)(M+4)},&\\
\langle a_{in}^2  a_{jm}^2  a_{kp}^2\rangle=\langle a_{in}^2 a_{im}^2 a_{kp}^2\rangle=
\langle a_{in}^2  a_{im}^2  a_{ip}^2\rangle=
\langle a_{in}^2  a_{jn}^2  a_{kp}^2\rangle=&\\ \langle a_{in}^2 a_{jn}^2 a_{ip}^2\rangle=
\langle a_{in}^2  a_{jn}^2  a_{jp}^2\rangle=\langle a_{in}^2 a_{jn}^2 a_{kn}^2\rangle=\df{1}{M(M+2)(M+4)}&.\\
\end{split} \]
This leads to the ensemble average:
 \begin{equation}
\langle\mbox{tr}\big(\rho_{12}^{T_2}\big)^3\rangle =\frac{N_1^2+N_2^2+N_3^2+3(N_1+N_2+N_3+N_1N_2N_3)}{(N_1N_2N_3+2)(N_1N_2N_3+4)},
\label{tr3real}
\end{equation}
where we have used $M=N_1N_2N_3$.

 Using the earlier statement that the averages prior to PT can be found from those after
 by the replacement $N_2 \rarrow 1$ and $N_1 \rarrow N=N_1N_2$, one gets the
\begin{equation}
\begin{split}
\langle\mbox{tr}\big(\rho_{12}\big)^3\rangle &=\frac{ N_1^2N_2^2+N_3^2+3\;N_1N_2 N_3+1}{(N_1N_2N_3+1)(N_1N_2N_3 +2)},\\ 
\langle\mbox{tr}\big(\rho_{12}\big)^3\rangle & =\frac{ N_1^2N_2^2+N_3^2+3\;(N_1N_2+N_3+N_1N_2N_3)+4}{(N_1N_2N_3+2)(N_1N_2N_3 +4)},
\end{split} 
\end{equation}
for the case of complex and real cases respectively. This indeed agrees with a previous calculation of this quantity, Eq.~(5.11) in \cite{Sommers04}, where the complex case
is considered.

\section{To show $\mbox{tr}\left(\rho_{12}^{T_2}\right)^m \neq \mbox{tr}\left(\rho_{23}^{T_3}\right)^m \neq 
\mbox{tr}\left(\rho_{31}^{T_1}\right)^m$ in general for $m>3$}
\label{highmoments}

In this Appendix it is shown that, in general, moments of  order higher than three of the density matrix after 
PT are not permutation symmetric. Using Eq.~(\ref{wavefunction1}) the following equation is obtained:
\beq
\label{trrhoPT3}
 \mbox{tr}\left(\rho_{12}^{T_2}\right)^3=\psi_{jk'l}\overline{\psi}^{j'kl} \psi_{j'k''l'}\overline{\psi}^{j''k'l'}
 \psi_{j''kl''} \overline{\psi}^{jk''l''}.
\eeq
Every index of a given tensor is contracted with a corresponding index in a  {\it distinct} dual tensor. 
This allows the Eq.~(\ref{trrhoPT3}) to be associated with a set of triples:
\beq
S_{12}=\{(b_2,b_1,b_0),(b_0,b_2,b_1),(b_1,b_0,b_2)\}. 
\eeq
This is to be understood as follows:
the dual tensors $\overline{\psi}$ that appear are labelled in their order or appearance from
left to right as $b_0$, $b_1$, $b_2$ and  the first triple $(b_2,b_1,b_0)$ indicates that the
first tensor $\psi$ is such that its first index is contracted with the third dual tensor, its second
with the second and the third with the first dual tensor. The second triple refers
to the contraction order for the second tensor and similarly the third. The association with the 
set of triples is not unique, for example any permutation among the $\{b_i\}$ and/or permutation
among the triples of the set refers to the same quantity. These operations correspond
to differently ordering the tensors and their duals.

%

Consider the following
\beq
 \mbox{tr}\left(\rho_{13}^{T_3}\right)^3=\psi_{jkl'}\overline{\psi}^{j'kl} \psi_{j'k'l''}\overline{\psi}^{j''k'l'} 
\psi_{j''k''l} \overline{\psi}^{jk''l''}.
\eeq
Following the above prescription allows to assign this quantity the set 
\beq
S_{13}=\{(b_2,b_0,b_1),(b_0,b_1,b_2),(b_1,b_2,b_0)\},
\eeq 
which is the same as the one for Eq.~(\ref{trrhoPT3}) if we interchange $b_1$ and $b_0$.

As a first case it is shown that fourth moment of density matrix after PT is not permutation
symmetric. This leads to the following sets of triples for  $\mbox{tr}\left(\rho_{12}^{T_2}\right)^4$ and 
$\mbox{tr}\left(\rho_{13}^{T_3}\right)^4$ respectively:
\beqa
S_{12}&=\{(b_3,b_1,b_0),(b_0,b_2,b_1),(b_1,b_3,b_2),(b_2,b_0,b_3)\}\\
S_{13}&=\{(b_3,b_0,b_1),(b_0,b_1,b_2),(b_1,b_2,b_3),(b_2,b_3,b_0)\}.
\eeqa
It is not hard to see that these two sets are not compatible under permutations of the $\{b_i\}$, and
hence the fourth moments are not the same. For example if we identify the first triples in the two 
sets, this implies that $b_0 \mapsto b_1$ and $b_1 \mapsto b_0$ (mapping direction being from $S_{13}$ to $S_{12}$).
This implies that the second triple of $S_{13}$ maps to $(b_1,b_0,-)$, where the $-$ indicates
some other $b_i$. However from $S_{12}$ it is seen that there are no triples that are like this. In fact any
identification of the triples leads to contradictions. 

On similar lines for $\mbox{tr}\left(\rho_{12}^{T_2}\right)^n$, $\mbox{tr}\left(\rho_{13}^{T_3}\right)^n$ it can be seen that the
associated sets are \beqa S_{12}&=\{ (b_{n-1} ,b_1 ,b_0), (b_0, b_2, b_1), (b_1, b_3, b_2),\ldots , 
(b_{n-3}, b_{n-1}, b_{n-2}), (b_{n-2}, b_0, b_{n-1})\}\\
S_{13}&=\{(b_{n-1}, b_0, b_1), (b_0, b_1, b_2), (b_1, b_2, b_3),\ldots, (b_{n-3}, b_{n-2},b_{n-1}), (b_{n-2}, b_{n-1}, b_0)\}.
\eeqa
Identify the any triple from $S_{12}$, $(b_p ,b_{p+2} ,b_{p+1})$, with any triple $(b_{r},b_{r+1},b_{r+2})$ from $S_{13}$, so that $b_r \mapsto b_p$, $b_{r+1} \mapsto b_{p+2}$, and $b_{r+2} \mapsto b_{p+1}$. It follows that the triple $(b_{r-1},b_r,b_{r+1})$ in $S_{13}$
maps to $(-, b_{p},b_{p+2})$. This can match with the corresponding term in $S_{12}$, with a $b_p$ at the center of the triple,
only if $( p-1) \,\mbox{mod}\,n \, =\, (p+2) \,\mbox{mod}\, n$, for all $0\le p \le n-1$, which implies that $n=1$ or $n=3$, these cases corresponding 
to the trivial $\mbox{tr}\left(\rho_{12}^{T_2}\right)=1$ and the non-trivial quantity $\mbox{tr}\left(\rho_{12}^{T_2}\right)^3$.

\section{Regarding the W-state example}
\label{app:counterexample}

In this appendix details of the example in Eq.~(\ref{Wstate}) with $\alpha^2=3/7$ and $\beta^2=\gamma^2=2/7$ is provided. Consider the generalized W-state $|\psi\rangle=\alpha|001\rangle+\beta|010\rangle+\gamma|100\rangle$
where $\alpha^2+\beta^2+\gamma^2=1$.
A straightforward calculation gives the eigenvalues of $\rho_{12}^{T_2}$ as $\{\beta^2$, $\gamma^2$, 
$\left(\alpha^2\pm\sqrt{\alpha^4+4\beta^2\gamma^2}\right)/2\}$, that of $\rho_{13}^{T_3}$ as $\{\alpha^2$, $\gamma^2$, 
$\left(\beta^2\pm\sqrt{\beta^4+4\alpha^2\gamma^2}\right)/2\}$ and that of $\rho_{23}^{T_3}$ as $\{\alpha^2$, $\beta^2, \left(\gamma^2\pm\sqrt{\gamma^4+4\alpha^2\beta^2}\right)/2\}$. Hence using these eigenvalues it immediately follows that
Eq.~(\ref{Winvariant}) holds for $\rho_{12}^{T_2}$, $\rho_{13}^{T_3}$ and $\rho_{23}^{T_3}$. 

Consider the case for which $\alpha=0$ and $|\psi\rangle=\left(\beta|01\rangle+\gamma|10\rangle\right)\otimes |0\rangle$, so that 
the first two qubits are entangled and they are in a product state with the third qubit. 
In this case the eigenvalues of $\rho_{12}^{T_2}$ are $\{\beta^2$, $\gamma^2$, $\pm\beta\gamma\}$
whereas the eigenvalues of $\rho_{13}^{T_3}$ and $\rho_{23}^{T_3}$ are $\{0, 0,\beta^2,\gamma^2\}$. Thus iff $n$ is odd the following holds:
\begin{equation}
\mbox{tr}\left(\rho_{12}^{T_2}\right)^n = \mbox{tr}\left(\rho_{23}^{T_3}\right)^n = 
\mbox{tr}\left(\rho_{13}^{T_3}\right)^n=\beta^{2n}+\gamma^{2n}.
\end{equation}
In general, it follows from considerations elaborated around Eq.~(\ref{eq:PTeigenvaluesPurestate}) that for {\it any} dimensional tripartite state with only two subsystems entangled, the odd moments of the density matrix
after PT are permutation symmetric whereas the even moments are not.

For special values of $\alpha$, $\beta$ and $\gamma$ it is shown that the moments of order higher than three of the
density matrix after PT are not permutation symmetric.
Special values that are considered here are 
$\alpha^2=3/7$, $\beta^2=\gamma^2=2/7$. In this case the eigenvalues of $\rho_{12}^{T_2}$ are $2/7$, $2/7$, $4/7$, $-1/7$
and that of $\rho_{13}^{T_3}$ are $3/7$, $2/7$, $\left(1+\sqrt{7}\right)/7$, $\left(1-\sqrt{7}\right)/7$. 
Thus $n$th moment of the density matrices after PT are given by Eq.~(\ref{nthmoment}).

Note that $\mbox{tr}\left(\rho_{13}^{T_3}\right)^n < \mbox{tr}\left(\rho_{12}^{T_2}\right)^n$ implies that 
$t_n=3^n+\left(1+\sqrt{7}\right)^n+\left(1-\sqrt{7}\right)^n < t_n'=2^n+4^n+(-1)^n$. The recursion  relations for $t_n$ and $t_n'$ are given by 
\beq t_{n+3}=5t_{n+2}-18t_n, \;\; t_{n+3}'=5t_{n+2}'-2t_{n+1}'-8t_n'\eeq
respectively. For example the first recursion relation is obtained by considering 3, $\left(1\pm\sqrt{7}\right)$ as roots of a cubic polynomial.
 The method of mathematical induction can be used to prove that indeed  $t_n < t_n'$ for $n>3$. Assume that
$t_n' > t_n$, $t_{n+1}' > t_{n+1}$, $t_{n+2}' > t_{n+2}$ which is true for $n=4$. then it is sufficient to show that
$t_{n+3}' > t_{n+3}$. It follows from the assumption that 
\begin{equation}
 t_{n+3}'=5t_{n+2}'-2t_{n+1}'-8t_n'>5t_{n+2}-2t_{n+1}'-8t_n'.
\end{equation}
Now if $5t_{n+2}-2t_{n+1}'-8t_n' > t_{n+3}$ then it follows that 
 $5t_{n+2}-2t_{n+1}'-8t_n' > 5t_{n+2}-18t_n$ or equivalently $18 t_n > 2 t_{n+1}'+8 t_n'$. However from the assumption $18 t_n' > 18 t_n$, and hence $18 t_n' >2 t_{n+1}'+8 t_n'$  which gives $5t_n' > t_{n+1}'$. Finally therefore 
$3(2^n)+ 4^n +6(-1)^n > 0$ which is certainly holds for $n>3$. Thus $t_{n+3}' > t_{n+3}$  for all $n>3$, as required to be proved.





\bibliography{ref2010}

\begin{thebibliography}{63}
\expandafter\ifx\csname natexlab\endcsname\relax\def\natexlab#1{#1}\fi
\expandafter\ifx\csname bibnamefont\endcsname\relax
  \def\bibnamefont#1{#1}\fi
\expandafter\ifx\csname bibfnamefont\endcsname\relax
  \def\bibfnamefont#1{#1}\fi
\expandafter\ifx\csname citenamefont\endcsname\relax
  \def\citenamefont#1{#1}\fi
\expandafter\ifx\csname url\endcsname\relax
  \def\url#1{\texttt{#1}}\fi
\expandafter\ifx\csname urlprefix\endcsname\relax\def\urlprefix{URL }\fi
\providecommand{\bibinfo}[2]{#2}
\providecommand{\eprint}[2][]{\url{#2}}

\bibitem[{\citenamefont{Einstien et~al.}(1935)\citenamefont{Einstien, Podolsky,
  and Rosen}}]{epr_paradox}
\bibinfo{author}{\bibfnamefont{A.}~\bibnamefont{Einstien}},
  \bibinfo{author}{\bibfnamefont{B.}~\bibnamefont{Podolsky}}, \bibnamefont{and}
  \bibinfo{author}{\bibfnamefont{N.}~\bibnamefont{Rosen}},
  \bibinfo{journal}{Phys. Rev.} \textbf{\bibinfo{volume}{47}},
  \bibinfo{pages}{777} (\bibinfo{year}{1935}).

\bibitem[{\citenamefont{Bell}(1964)}]{bell}
\bibinfo{author}{\bibfnamefont{J.}~\bibnamefont{Bell}},
  \bibinfo{journal}{Physics} \textbf{\bibinfo{volume}{1}}, \bibinfo{pages}{195}
  (\bibinfo{year}{1964}).

\bibitem[{\citenamefont{Aspect et~al.}(1982)\citenamefont{Aspect, Dalibard, and
  Roger}}]{aspect}
\bibinfo{author}{\bibfnamefont{A.}~\bibnamefont{Aspect}},
  \bibinfo{author}{\bibnamefont{Dalibard}}, \bibnamefont{and}
  \bibinfo{author}{\bibfnamefont{G.}~\bibnamefont{Roger}},
  \bibinfo{journal}{Phys. Rev. Lett.} \textbf{\bibinfo{volume}{49}},
  \bibinfo{pages}{1804} (\bibinfo{year}{1982}).

\bibitem[{\citenamefont{Jozsa and Linden}(2003)}]{Jozsalinden}
\bibinfo{author}{\bibfnamefont{R.}~\bibnamefont{Jozsa}} \bibnamefont{and}
  \bibinfo{author}{\bibfnamefont{N.}~\bibnamefont{Linden}},
  \bibinfo{journal}{Proc. R. Soc. A} \textbf{\bibinfo{volume}{459}},
  \bibinfo{pages}{2011} (\bibinfo{year}{2003}).

\bibitem[{\citenamefont{Bennett et~al.}(1993)\citenamefont{Bennett, Brassard,
  Crepeau, Jozsa, Peres, and Wootters}}]{Teleport}
\bibinfo{author}{\bibfnamefont{C.~H.} \bibnamefont{Bennett}},
  \bibinfo{author}{\bibfnamefont{G.}~\bibnamefont{Brassard}},
  \bibinfo{author}{\bibfnamefont{C.}~\bibnamefont{Crepeau}},
  \bibinfo{author}{\bibfnamefont{R.}~\bibnamefont{Jozsa}},
  \bibinfo{author}{\bibfnamefont{A.}~\bibnamefont{Peres}}, \bibnamefont{and}
  \bibinfo{author}{\bibfnamefont{W.}~\bibnamefont{Wootters}},
  \bibinfo{journal}{Phys. Rev. Lett.} \textbf{\bibinfo{volume}{70}},
  \bibinfo{pages}{1895} (\bibinfo{year}{1993}).

\bibitem[{\citenamefont{Bennett and Wiesner}(1992)}]{Superdense}
\bibinfo{author}{\bibfnamefont{C.~H.} \bibnamefont{Bennett}} \bibnamefont{and}
  \bibinfo{author}{\bibfnamefont{S.~J.} \bibnamefont{Wiesner}},
  \bibinfo{journal}{Phys. Rev. Lett.} \textbf{\bibinfo{volume}{69}},
  \bibinfo{pages}{2881} (\bibinfo{year}{1992}).

\bibitem[{\citenamefont{Masanes}(2006)}]{Masanes}
\bibinfo{author}{\bibfnamefont{L.}~\bibnamefont{Masanes}},
  \bibinfo{journal}{Phys. Rev. Lett.} \textbf{\bibinfo{volume}{96}},
  \bibinfo{pages}{150501} (\bibinfo{year}{2006}).

\bibitem[{\citenamefont{Piani and Watrous}(2009)}]{Pianiwatrous}
\bibinfo{author}{\bibfnamefont{M.}~\bibnamefont{Piani}} \bibnamefont{and}
  \bibinfo{author}{\bibfnamefont{J.}~\bibnamefont{Watrous}},
  \bibinfo{journal}{Phys. Rev. Lett.} \textbf{\bibinfo{volume}{102}},
  \bibinfo{pages}{250501} (\bibinfo{year}{2009}).

\bibitem[{\citenamefont{Ghosh et~al.}(2003)\citenamefont{Ghosh, Rosenbaum,
  Aeppli, and Coppersmith}}]{Ghosh}
\bibinfo{author}{\bibfnamefont{S.}~\bibnamefont{Ghosh}},
  \bibinfo{author}{\bibfnamefont{T.~F.} \bibnamefont{Rosenbaum}},
  \bibinfo{author}{\bibfnamefont{G.}~\bibnamefont{Aeppli}}, \bibnamefont{and}
  \bibinfo{author}{\bibfnamefont{S.~N.} \bibnamefont{Coppersmith}},
  \bibinfo{journal}{Nature} \textbf{\bibinfo{volume}{425}}, \bibinfo{pages}{48}
  (\bibinfo{year}{2003}).

\bibitem[{\citenamefont{Lubkin}(1978)}]{Lubkin}
\bibinfo{author}{\bibfnamefont{E.}~\bibnamefont{Lubkin}}, \bibinfo{journal}{J.
  Math. Phys.} \textbf{\bibinfo{volume}{19}}, \bibinfo{pages}{1028}
  (\bibinfo{year}{1978}).

\bibitem[{\citenamefont{Page}(1993)}]{Page}
\bibinfo{author}{\bibfnamefont{D.}~\bibnamefont{Page}}, \bibinfo{journal}{Phys.
  Rev. Lett.} \textbf{\bibinfo{volume}{71}}, \bibinfo{pages}{9}
  (\bibinfo{year}{1993}).

\bibitem[{\citenamefont{Hayden et~al.}(2006)\citenamefont{Hayden, Leung, and
  Winter}}]{Haydenmath}
\bibinfo{author}{\bibfnamefont{P.}~\bibnamefont{Hayden}},
  \bibinfo{author}{\bibfnamefont{D.~W.} \bibnamefont{Leung}}, \bibnamefont{and}
  \bibinfo{author}{\bibfnamefont{A.}~\bibnamefont{Winter}},
  \bibinfo{journal}{Commun. Math. Phys.} \textbf{\bibinfo{volume}{265}},
  \bibinfo{pages}{95} (\bibinfo{year}{2006}).

\bibitem[{\citenamefont{Bandyopadhyay and Lakshminarayan}(2002)}]{Arul}
\bibinfo{author}{\bibfnamefont{J.~N.} \bibnamefont{Bandyopadhyay}}
  \bibnamefont{and}
  \bibinfo{author}{\bibfnamefont{A.}~\bibnamefont{Lakshminarayan}},
  \bibinfo{journal}{Phys. Rev. Lett.} \textbf{\bibinfo{volume}{89}},
  \bibinfo{pages}{060402} (\bibinfo{year}{2002}).

\bibitem[{\citenamefont{Haake}(2010)}]{HaakeBook}
\bibinfo{author}{\bibfnamefont{F.}~\bibnamefont{Haake}},
  \emph{\bibinfo{title}{Quantum Signatures of Chaos}}
  (\bibinfo{publisher}{Springer, 3rd Edition, Berlin}, \bibinfo{year}{2010}).

\bibitem[{\citenamefont{Beenakker}(1997)}]{beenakker97}
\bibinfo{author}{\bibfnamefont{C.~W.~J.} \bibnamefont{Beenakker}},
  \bibinfo{journal}{Rev. Mod. Phys.} \textbf{\bibinfo{volume}{69}},
  \bibinfo{pages}{731–808} (\bibinfo{year}{1997}).

\bibitem[{\citenamefont{Brody et~al.}(1981)\citenamefont{Brody, Flores, French,
  Mello, Pandey, and Wong}}]{brody81}
\bibinfo{author}{\bibfnamefont{T.~A.} \bibnamefont{Brody}},
  \bibinfo{author}{\bibfnamefont{J.}~\bibnamefont{Flores}},
  \bibinfo{author}{\bibfnamefont{J.~B.} \bibnamefont{French}},
  \bibinfo{author}{\bibfnamefont{P.~A.} \bibnamefont{Mello}},
  \bibinfo{author}{\bibfnamefont{A.}~\bibnamefont{Pandey}}, \bibnamefont{and}
  \bibinfo{author}{\bibfnamefont{S.~S.~M.} \bibnamefont{Wong}},
  \bibinfo{journal}{Rev. Mod. Phys.} \textbf{\bibinfo{volume}{53}},
  \bibinfo{pages}{385} (\bibinfo{year}{1981}).

\bibitem[{\citenamefont{Peres}(1996)}]{Peres}
\bibinfo{author}{\bibfnamefont{A.}~\bibnamefont{Peres}},
  \bibinfo{journal}{Phys. Rev. Lett.} \textbf{\bibinfo{volume}{77}},
  \bibinfo{pages}{1413–1415} (\bibinfo{year}{1996}).

\bibitem[{\citenamefont{Horodecki et~al.}(1996)\citenamefont{Horodecki,
  Horodecki, and Horodecki}}]{mhorodecki}
\bibinfo{author}{\bibfnamefont{M.}~\bibnamefont{Horodecki}},
  \bibinfo{author}{\bibfnamefont{P.}~\bibnamefont{Horodecki}},
  \bibnamefont{and}
  \bibinfo{author}{\bibfnamefont{R.}~\bibnamefont{Horodecki}},
  \bibinfo{journal}{Physics Letters A} \textbf{\bibinfo{volume}{223}},
  \bibinfo{pages}{1} (\bibinfo{year}{1996}).

\bibitem[{\citenamefont{Hill and Wootters}(1997)}]{Wooters}
\bibinfo{author}{\bibfnamefont{S.}~\bibnamefont{Hill}} \bibnamefont{and}
  \bibinfo{author}{\bibfnamefont{W.~K.} \bibnamefont{Wootters}},
  \bibinfo{journal}{Phys. Rev. Lett.} \textbf{\bibinfo{volume}{78}},
  \bibinfo{pages}{5022} (\bibinfo{year}{1997}).

\bibitem[{\citenamefont{Wootters}(1998)}]{Wootersentform}
\bibinfo{author}{\bibfnamefont{W.~K.} \bibnamefont{Wootters}},
  \bibinfo{journal}{Phys. Rev. Lett.} \textbf{\bibinfo{volume}{80}},
  \bibinfo{pages}{10} (\bibinfo{year}{1998}).

\bibitem[{\citenamefont{Vidal and Werner}(2002)}]{vidal}
\bibinfo{author}{\bibfnamefont{G.}~\bibnamefont{Vidal}} \bibnamefont{and}
  \bibinfo{author}{\bibfnamefont{R.~F.} \bibnamefont{Werner}},
  \bibinfo{journal}{Phys. Rev. A} \textbf{\bibinfo{volume}{65}},
  \bibinfo{pages}{032314} (\bibinfo{year}{2002}).

\bibitem[{\citenamefont{Plenio}(2005)}]{logneg}
\bibinfo{author}{\bibfnamefont{M.~B.} \bibnamefont{Plenio}},
  \bibinfo{journal}{Phys. Rev. Lett.} \textbf{\bibinfo{volume}{95}},
  \bibinfo{pages}{090503} (\bibinfo{year}{2005}).

\bibitem[{\citenamefont{Horodecki et~al.}(1998)\citenamefont{Horodecki,
  Horodecki, and Horodecki}}]{mhorodeckibound}
\bibinfo{author}{\bibfnamefont{M.}~\bibnamefont{Horodecki}},
  \bibinfo{author}{\bibfnamefont{P.}~\bibnamefont{Horodecki}},
  \bibnamefont{and}
  \bibinfo{author}{\bibfnamefont{R.}~\bibnamefont{Horodecki}},
  \bibinfo{journal}{Phys. Rev. Lett.} \textbf{\bibinfo{volume}{80}},
  \bibinfo{pages}{5239} (\bibinfo{year}{1998}).

\bibitem[{\citenamefont{Kempe}(1999)}]{Kempe99}
\bibinfo{author}{\bibfnamefont{J.}~\bibnamefont{Kempe}},
  \bibinfo{journal}{Phys. Rev. A} \textbf{\bibinfo{volume}{60}},
  \bibinfo{pages}{910} (\bibinfo{year}{1999}).

\bibitem[{\citenamefont{Bohigas et~al.}(1984)\citenamefont{Bohigas, Giannoni,
  and Schmit}}]{Bohigas84}
\bibinfo{author}{\bibfnamefont{O.}~\bibnamefont{Bohigas}},
  \bibinfo{author}{\bibfnamefont{M.~J.} \bibnamefont{Giannoni}},
  \bibnamefont{and} \bibinfo{author}{\bibfnamefont{C.}~\bibnamefont{Schmit}},
  \bibinfo{journal}{Phys. Rev. Lett.} \textbf{\bibinfo{volume}{52}},
  \bibinfo{pages}{1} (\bibinfo{year}{1984}).

\bibitem[{\citenamefont{Datta}(2010)}]{Animeshrandom}
\bibinfo{author}{\bibfnamefont{A.}~\bibnamefont{Datta}},
  \bibinfo{journal}{Phys. Rev. A} \textbf{\bibinfo{volume}{81}},
  \bibinfo{pages}{052312} (\bibinfo{year}{2010}).

\bibitem[{\citenamefont{Nadal et~al.}(2011)\citenamefont{Nadal, Majumdar, and
  Vergassola}}]{Nadal11}
\bibinfo{author}{\bibfnamefont{C.}~\bibnamefont{Nadal}},
  \bibinfo{author}{\bibfnamefont{S.~N.} \bibnamefont{Majumdar}},
  \bibnamefont{and}
  \bibinfo{author}{\bibfnamefont{M.}~\bibnamefont{Vergassola}},
  \bibinfo{journal}{J. Stat. Phys.} \textbf{\bibinfo{volume}{142}},
  \bibinfo{pages}{403} (\bibinfo{year}{2011}).

\bibitem[{\citenamefont{Znidaric et~al.}(2007)\citenamefont{Znidaric, Prosen,
  Benenti, and Casati}}]{Marko}
\bibinfo{author}{\bibfnamefont{M.}~\bibnamefont{Znidaric}},
  \bibinfo{author}{\bibfnamefont{T.}~\bibnamefont{Prosen}},
  \bibinfo{author}{\bibfnamefont{G.}~\bibnamefont{Benenti}}, \bibnamefont{and}
  \bibinfo{author}{\bibfnamefont{G.}~\bibnamefont{Casati}},
  \bibinfo{journal}{J. Phys. A: Math. Theor.} \textbf{\bibinfo{volume}{40}},
  \bibinfo{pages}{13787} (\bibinfo{year}{2007}).

\bibitem[{\citenamefont{Kendon et~al.}(2002{\natexlab{a}})\citenamefont{Kendon,
  Zyczkowski, and Munro}}]{Kendonpra}
\bibinfo{author}{\bibfnamefont{V.~M.} \bibnamefont{Kendon}},
  \bibinfo{author}{\bibfnamefont{K.}~\bibnamefont{Zyczkowski}},
  \bibnamefont{and} \bibinfo{author}{\bibfnamefont{W.~J.} \bibnamefont{Munro}},
  \bibinfo{journal}{Phys. Rev. A} \textbf{\bibinfo{volume}{66}},
  \bibinfo{pages}{062310} (\bibinfo{year}{2002}{\natexlab{a}}).

\bibitem[{\citenamefont{Kendon et~al.}(2002{\natexlab{b}})\citenamefont{Kendon,
  K.Nemoto, and Munro}}]{Kendonopt}
\bibinfo{author}{\bibfnamefont{V.~M.} \bibnamefont{Kendon}},
  \bibinfo{author}{\bibnamefont{K.Nemoto}}, \bibnamefont{and}
  \bibinfo{author}{\bibfnamefont{W.~J.} \bibnamefont{Munro}},
  \bibinfo{journal}{J. Mod. Optics} \textbf{\bibinfo{volume}{49}},
  \bibinfo{pages}{1709} (\bibinfo{year}{2002}{\natexlab{b}}).

\bibitem[{\citenamefont{Carteret}(2005)}]{Hilarycircuit}
\bibinfo{author}{\bibfnamefont{H.~A.} \bibnamefont{Carteret}},
  \bibinfo{journal}{Phys. Rev. Lett.} \textbf{\bibinfo{volume}{94}},
  \bibinfo{pages}{040502} (\bibinfo{year}{2005}).

\bibitem[{\citenamefont{Sommers and Zyczkowski}(2004)}]{Sommers04}
\bibinfo{author}{\bibfnamefont{H.-J.} \bibnamefont{Sommers}} \bibnamefont{and}
  \bibinfo{author}{\bibfnamefont{K.}~\bibnamefont{Zyczkowski}},
  \bibinfo{journal}{J. Phys. A: Math. Gen.} \textbf{\bibinfo{volume}{37}},
  \bibinfo{pages}{8457} (\bibinfo{year}{2004}).

\bibitem[{\citenamefont{Borot and Nadal}()}]{Borot11}
\bibinfo{author}{\bibfnamefont{G.}~\bibnamefont{Borot}} \bibnamefont{and}
  \bibinfo{author}{\bibfnamefont{C.}~\bibnamefont{Nadal}},
  \bibinfo{note}{arXiv:1110.3838}.

\bibitem[{\citenamefont{O.~Giraud and Georgeot}(2007)}]{Giraud07}
\bibinfo{author}{\bibfnamefont{J.~M.} \bibnamefont{O.~Giraud}}
  \bibnamefont{and} \bibinfo{author}{\bibfnamefont{B.}~\bibnamefont{Georgeot}},
  \bibinfo{journal}{Phys. Rev. A} \textbf{\bibinfo{volume}{76}},
  \bibinfo{pages}{042333} (\bibinfo{year}{2007}).

\bibitem[{\citenamefont{Meyer and Wallach}(2002)}]{Meyer02}
\bibinfo{author}{\bibfnamefont{D.~A.} \bibnamefont{Meyer}} \bibnamefont{and}
  \bibinfo{author}{\bibfnamefont{N.~R.} \bibnamefont{Wallach}},
  \bibinfo{journal}{J. Math. Phys} \textbf{\bibinfo{volume}{43}},
  \bibinfo{pages}{4273} (\bibinfo{year}{2002}).

\bibitem[{\citenamefont{O.~Giraud and Georgeot}(2009)}]{Giraud09}
\bibinfo{author}{\bibfnamefont{J.~M.} \bibnamefont{O.~Giraud}}
  \bibnamefont{and} \bibinfo{author}{\bibfnamefont{B.}~\bibnamefont{Georgeot}},
  \bibinfo{journal}{Phys. Rev. A} \textbf{\bibinfo{volume}{79}},
  \bibinfo{pages}{032308} (\bibinfo{year}{2009}).

\bibitem[{\citenamefont{Aubrun}()}]{Aubrun10}
\bibinfo{author}{\bibfnamefont{G.}~\bibnamefont{Aubrun}},
  \bibinfo{note}{arXiv:1011.0275v2 [math.PR]}.

\bibitem[{\citenamefont{Benoit~Collins}()}]{Collins11}
\bibinfo{author}{\bibfnamefont{D.~Y.} \bibnamefont{Benoit~Collins},
  \bibfnamefont{Ion~Nechita}}, \bibinfo{note}{arXiv:1108.1935v1 [math.ph]}.

\bibitem[{\citenamefont{Lloyd and Pagels}(1988)}]{llyod}
\bibinfo{author}{\bibfnamefont{S.}~\bibnamefont{Lloyd}} \bibnamefont{and}
  \bibinfo{author}{\bibfnamefont{H.}~\bibnamefont{Pagels}},
  \bibinfo{journal}{Ann. Phys.} \textbf{\bibinfo{volume}{188}},
  \bibinfo{pages}{186} (\bibinfo{year}{1988}).

\bibitem[{\citenamefont{Zyczkowski and Sommers}(2001)}]{Sommers}
\bibinfo{author}{\bibfnamefont{K.}~\bibnamefont{Zyczkowski}} \bibnamefont{and}
  \bibinfo{author}{\bibfnamefont{H.-J.} \bibnamefont{Sommers}},
  \bibinfo{journal}{J. Phys. A: Math. Gen.} \textbf{\bibinfo{volume}{34}},
  \bibinfo{pages}{7111} (\bibinfo{year}{2001}).

\bibitem[{\citenamefont{Marcenko and Pastur}(1967)}]{Marcenko}
\bibinfo{author}{\bibfnamefont{V.}~\bibnamefont{Marcenko}} \bibnamefont{and}
  \bibinfo{author}{\bibfnamefont{L.}~\bibnamefont{Pastur}},
  \bibinfo{journal}{Math. USSR-Sb} \textbf{\bibinfo{volume}{1}},
  \bibinfo{pages}{457} (\bibinfo{year}{1967}).

\bibitem[{\citenamefont{Sen}(1996)}]{Sen}
\bibinfo{author}{\bibfnamefont{S.}~\bibnamefont{Sen}}, \bibinfo{journal}{Phys.
  Rev. Lett.} \textbf{\bibinfo{volume}{77}}, \bibinfo{pages}{1}
  (\bibinfo{year}{1996}).

\bibitem[{\citenamefont{Sanchez-Ruiz}(1995)}]{Jorge}
\bibinfo{author}{\bibfnamefont{J.}~\bibnamefont{Sanchez-Ruiz}},
  \bibinfo{journal}{Phys. Rev. E} \textbf{\bibinfo{volume}{52}},
  \bibinfo{pages}{5653} (\bibinfo{year}{1995}).

\bibitem[{\citenamefont{Mehta}(2004)}]{Mehtabook}
\bibinfo{author}{\bibfnamefont{M.~L.} \bibnamefont{Mehta}},
  \emph{\bibinfo{title}{Random Matrices}} (\bibinfo{publisher}{Elsevier
  Academic Press, 3rd Edition, London}, \bibinfo{year}{2004}).

\bibitem[{\citenamefont{Forrester}(2010)}]{Forresterbook}
\bibinfo{author}{\bibfnamefont{P.~J.} \bibnamefont{Forrester}},
  \emph{\bibinfo{title}{Log-Gases and Random Matrices}}
  (\bibinfo{publisher}{Princeton University Press, Princeton and Oxford},
  \bibinfo{year}{2010}).

\bibitem[{\citenamefont{Vijayaraghavan
  et~al.}(2011)\citenamefont{Vijayaraghavan, T.Bhosale, and
  Lakshminarayan}}]{Vikram11}
\bibinfo{author}{\bibfnamefont{V.~S.} \bibnamefont{Vijayaraghavan}},
  \bibinfo{author}{\bibfnamefont{U.}~\bibnamefont{T.Bhosale}},
  \bibnamefont{and}
  \bibinfo{author}{\bibfnamefont{A.}~\bibnamefont{Lakshminarayan}},
  \bibinfo{journal}{Phys. Rev. A} \textbf{\bibinfo{volume}{84}},
  \bibinfo{pages}{032306} (\bibinfo{year}{2011}).

\bibitem[{\citenamefont{Sudbery}(2001)}]{Sudbery01}
\bibinfo{author}{\bibfnamefont{A.}~\bibnamefont{Sudbery}}, \bibinfo{journal}{J.
  Phys. A: Math. Gen.} \textbf{\bibinfo{volume}{34}}, \bibinfo{pages}{643}
  (\bibinfo{year}{2001}).

\bibitem[{\citenamefont{Williamson et~al.}(2011)\citenamefont{Williamson,
  Ericsson, Johansson, Sj\"oqvist, Sudbery, Vedral, and
  Wootters}}]{Williamson11}
\bibinfo{author}{\bibfnamefont{M.~S.} \bibnamefont{Williamson}},
  \bibinfo{author}{\bibfnamefont{M.}~\bibnamefont{Ericsson}},
  \bibinfo{author}{\bibfnamefont{M.}~\bibnamefont{Johansson}},
  \bibinfo{author}{\bibfnamefont{E.}~\bibnamefont{Sj\"oqvist}},
  \bibinfo{author}{\bibfnamefont{A.}~\bibnamefont{Sudbery}},
  \bibinfo{author}{\bibfnamefont{V.}~\bibnamefont{Vedral}}, \bibnamefont{and}
  \bibinfo{author}{\bibfnamefont{W.~K.} \bibnamefont{Wootters}},
  \bibinfo{journal}{Phys. Rev. A} \textbf{\bibinfo{volume}{83}},
  \bibinfo{pages}{062308} (\bibinfo{year}{2011}).

\bibitem[{\citenamefont{Ullah}(1983)}]{Ullah83}
\bibinfo{author}{\bibfnamefont{N.}~\bibnamefont{Ullah}}, \bibinfo{journal}{J.
  Phys. A: Math. Gen.} \textbf{\bibinfo{volume}{16}}, \bibinfo{pages}{L767}
  (\bibinfo{year}{1983}).

\bibitem[{\citenamefont{Tracy and Widom}(1994)}]{Tracy1}
\bibinfo{author}{\bibfnamefont{C.}~\bibnamefont{Tracy}} \bibnamefont{and}
  \bibinfo{author}{\bibfnamefont{H.}~\bibnamefont{Widom}},
  \bibinfo{journal}{Commun. Math. Phys.} \textbf{\bibinfo{volume}{159}},
  \bibinfo{pages}{151} (\bibinfo{year}{1994}).

\bibitem[{\citenamefont{Tracy and Widom}(1996)}]{Tracy2}
\bibinfo{author}{\bibfnamefont{C.}~\bibnamefont{Tracy}} \bibnamefont{and}
  \bibinfo{author}{\bibfnamefont{H.}~\bibnamefont{Widom}},
  \bibinfo{journal}{Commun. Math. Phys.} \textbf{\bibinfo{volume}{177}},
  \bibinfo{pages}{727} (\bibinfo{year}{1996}).

\bibitem[{\citenamefont{Edelman and Persson}()}]{Edelman05}
\bibinfo{author}{\bibfnamefont{A.}~\bibnamefont{Edelman}} \bibnamefont{and}
  \bibinfo{author}{\bibfnamefont{P.-O.} \bibnamefont{Persson}},
  \bibinfo{note}{arXiv:math-ph/0501068v1}.

\bibitem[{\citenamefont{Edelman and Rao}(2005)}]{Edelacta05}
\bibinfo{author}{\bibfnamefont{A.}~\bibnamefont{Edelman}} \bibnamefont{and}
  \bibinfo{author}{\bibfnamefont{N.~R.} \bibnamefont{Rao}},
  \bibinfo{journal}{Acta Numerica} \textbf{\bibinfo{volume}{14}},
  \bibinfo{pages}{233} (\bibinfo{year}{2005}).

\bibitem[{\citenamefont{Casati and Chirikov}(1995)}]{Chirikovbook}
\bibinfo{author}{\bibfnamefont{G.}~\bibnamefont{Casati}} \bibnamefont{and}
  \bibinfo{author}{\bibfnamefont{B.~V.} \bibnamefont{Chirikov}},
  \emph{\bibinfo{title}{Quantum chaos: between order and disorder : a selection
  of papers}} (\bibinfo{publisher}{Cambridge University Press},
  \bibinfo{year}{1995}).

\bibitem[{\citenamefont{Casati and Ford}(1979)}]{CasatiFord}
\bibinfo{author}{\bibfnamefont{G.}~\bibnamefont{Casati}} \bibnamefont{and}
  \bibinfo{author}{\bibfnamefont{J.}~\bibnamefont{Ford}},
  \emph{\bibinfo{title}{Stochastic Behavior In Classical and Quantum
  Hamiltonian Systems : Volta Memorial Conference, Como 1977}}
  (\bibinfo{publisher}{Springer}, \bibinfo{year}{1979}).

\bibitem[{\citenamefont{Izrailev}(1990)}]{Izrailev90}
\bibinfo{author}{\bibfnamefont{F.~M.} \bibnamefont{Izrailev}},
  \bibinfo{journal}{Physics Reports} \textbf{\bibinfo{volume}{196}},
  \bibinfo{pages}{299} (\bibinfo{year}{1990}).

\bibitem[{\citenamefont{Lakshminarayan}(2001)}]{Arulentpow}
\bibinfo{author}{\bibfnamefont{A.}~\bibnamefont{Lakshminarayan}},
  \bibinfo{journal}{Phys. Rev. E} \textbf{\bibinfo{volume}{64}},
  \bibinfo{pages}{036207} (\bibinfo{year}{2001}).

\bibitem[{\citenamefont{Wang and Garcia-Garcia}(2009)}]{Wang3drotor}
\bibinfo{author}{\bibfnamefont{J.}~\bibnamefont{Wang}} \bibnamefont{and}
  \bibinfo{author}{\bibfnamefont{A.~M.} \bibnamefont{Garcia-Garcia}},
  \bibinfo{journal}{Phys. Rev. E} \textbf{\bibinfo{volume}{79}},
  \bibinfo{pages}{036206} (\bibinfo{year}{2009}).

\bibitem[{\citenamefont{Lakshminarayan and Tomsovic}(2011)}]{Lakshminarayan}
\bibinfo{author}{\bibfnamefont{A.}~\bibnamefont{Lakshminarayan}}
  \bibnamefont{and} \bibinfo{author}{\bibfnamefont{S.}~\bibnamefont{Tomsovic}},
  \bibinfo{journal}{Phys.~Rev.~E} \textbf{\bibinfo{volume}{84}},
  \bibinfo{pages}{016218} (\bibinfo{year}{2011}).

\bibitem[{\citenamefont{Lichtenberg and Lieberman}(1992)}]{Liebermanbook}
\bibinfo{author}{\bibfnamefont{A.~J.} \bibnamefont{Lichtenberg}}
  \bibnamefont{and} \bibinfo{author}{\bibfnamefont{M.~A.}
  \bibnamefont{Lieberman}}, \emph{\bibinfo{title}{Regular and Chaotic
  Dynamics}} (\bibinfo{publisher}{Springer-Verlag, New York},
  \bibinfo{year}{1992}).

\bibitem[{\citenamefont{Reichl}(2004)}]{Reichlbook}
\bibinfo{author}{\bibfnamefont{L.~E.} \bibnamefont{Reichl}},
  \emph{\bibinfo{title}{The Transition to Chaos, 2nd edition}}
  (\bibinfo{publisher}{Springer-Verlag, New York}, \bibinfo{year}{2004}).

\bibitem[{\citenamefont{Lakshminarayan}(1997)}]{Arul97}
\bibinfo{author}{\bibfnamefont{A.}~\bibnamefont{Lakshminarayan}},
  \bibinfo{journal}{Pramana} \textbf{\bibinfo{volume}{48}},
  \bibinfo{pages}{517} (\bibinfo{year}{1997}).

\bibitem[{\citenamefont{Ullah and Porter}(1963)}]{Ullah}
\bibinfo{author}{\bibfnamefont{N.}~\bibnamefont{Ullah}} \bibnamefont{and}
  \bibinfo{author}{\bibfnamefont{C.~E.} \bibnamefont{Porter}},
  \bibinfo{journal}{Phys. Lett.} \textbf{\bibinfo{volume}{6}},
  \bibinfo{pages}{301} (\bibinfo{year}{1963}).

\end{thebibliography}

\end{document}